\documentclass[aps,prl,twocolumn,showpacs,superscriptaddress]{revtex4-2}  
\usepackage{graphicx}  
\usepackage{bm}        
\usepackage{amssymb}   
\usepackage{amsmath}   
\usepackage[colorlinks=true,allcolors=blue,breaklinks]{hyperref}
\usepackage{upgreek}
\usepackage{chemmacros}[2014/01/24]
\usepackage{color}

\newcommand{\be}{\begin{equation}}
\newcommand{\ft}[1]{{\cal #1}}
\newcommand{\ee}{\end{equation}}
\newcommand{\bea}{\begin{eqnarray}}
\newcommand{\eea}{\end{eqnarray}}

\newcommand{\la}{\langle}
\newcommand{\ra}{\rangle}

\newcommand{\lb}{\left[}
\newcommand{\rb}{\right]}
\newcommand{\lp}{\left(}
\newcommand{\rp}{\right)}

\renewcommand{\vec}[1]{{\bf #1}}
\newcommand{\eps}{\varepsilon}
\renewcommand{\tilde}{\widetilde}
\newcommand{\zero}{{(0)}}
\newcommand{\one}{{\rm a}}
\newcommand{\two}{{\rm p}}

\def\nn{\nonumber\\}

\graphicspath{{figures/}}

\def\afflux{Department of Physics and Material Science, University of Luxembourg, Luxembourg}
\def\affind{Research Center for Quantum Physics, National Research and Innovation Agency, South Tangerang, Indonesia}

\begin{document}
\title{Coulomb drag of viscous electron fluids:\\ drag viscosity and negative drag conductivity}

\author{Eddwi H. Hasdeo}
\affiliation{\afflux}
\affiliation{\affind}

\author{Edvin G. Idrisov}
\affiliation{\afflux}

\author{Thomas L. Schmidt}
\affiliation{\afflux}
\date{\today}

\begin{abstract}
We show that Coulomb drag in hydrodynamic bilayer systems leads to additional viscosity terms in the hydrodynamic equations, i.e., the drag and drag-Hall viscosities, besides the well-known kinematic and Hall viscosities.
These new viscosity terms arise from a change of the stress tensor due to the interlayer Coulomb interactions. All four viscosity terms are tunable by varying the applied magnetic field and the electron densities in the two layers. At certain ratios between the electron densities in the two layers, the drag viscosity dramatically changes the longitudinal transport resulting in a negative drag conductivity.
\end{abstract}

\maketitle

Several decades ago, Gurzhi imagined an ideal metal from which all impurities and scatterers (e.g.~phonons) were removed and which contained only electrons interacting among themselves~\cite{gurzhi1963}. In this case, the electrons behave collectively like a viscous fluid with a re\-sis\-ti\-vity determined by their viscosity, which is inversely proportional to temperature~\cite{andreev11-prl}. This result differs starkly from that in a normal metal whose resistivity increases with temperature due to electron-phonon interactions. Such hydrodynamic electron flows have been realized in clean samples of GaAs~\cite{DeJong1995}, graphene~\cite{Crossno2016,bandurin16-hydro,Levitov2016}, PdCoO$_2$~\cite{moll16-pdco-visc}, and in Weyl semimetals~\cite{Gooth2018}.

The viscous hydrodynamic regime can host many surprising transport phenomena, such as for instance an increase of the thermal conductivity and a breakdown of the Wiedemann-Franz law in graphene~\cite{Crossno2016}, an increase of the electrical conductance of a graphene constriction due to superballistic behavior of viscous flow~\cite{Guo3068,KrishnaKumar2017},  a nonlocal negative resistance in graphene~\cite{Levitov2016}, as well as peculiar electron flow in topological materials~\cite{hasdeo21-hydro, toshio21-hydro}. As the viscosity plays a central role in the transport of viscous electrons, it is natural to ask how can we manipulate the viscosity. It is well-known that to a certain extent the viscosity can be controlled by varying the temperature, the carrier density, and the impurity concentration~\cite{pellegrino17-hallvisc}.  On the other hand, applying a magnetic field not only modifies the viscosity but also introduces an additional \emph{Hall} viscosity in the hydrodynamic equations \cite{alekseev16-hallvisc,pellegrino17-hallvisc,scaffidi17-hallvisc,berdyugin19-hallvisc}.

As electrons are charged particles, one can place two layered metals parallel to each other and the interlayer Coulomb interaction will induce a drag voltage in the ``passive'' layer due to an applied current in the ``active'' layer~\cite{boris-cd}. If we consider the two metals in such a Coulomb drag setup as viscous fluids, we can ask if the viscosities of the two metals are modified due to the interlayer Coulomb interaction~\cite{galitski20-dragvis}. Furthermore, one could expect that the hydrodynamic equations might be modified because additional viscosity terms emerge from the interlayer Coulomb interaction similar to the case of Hall viscosity.

\begin{figure}
\includegraphics[width=6cm]{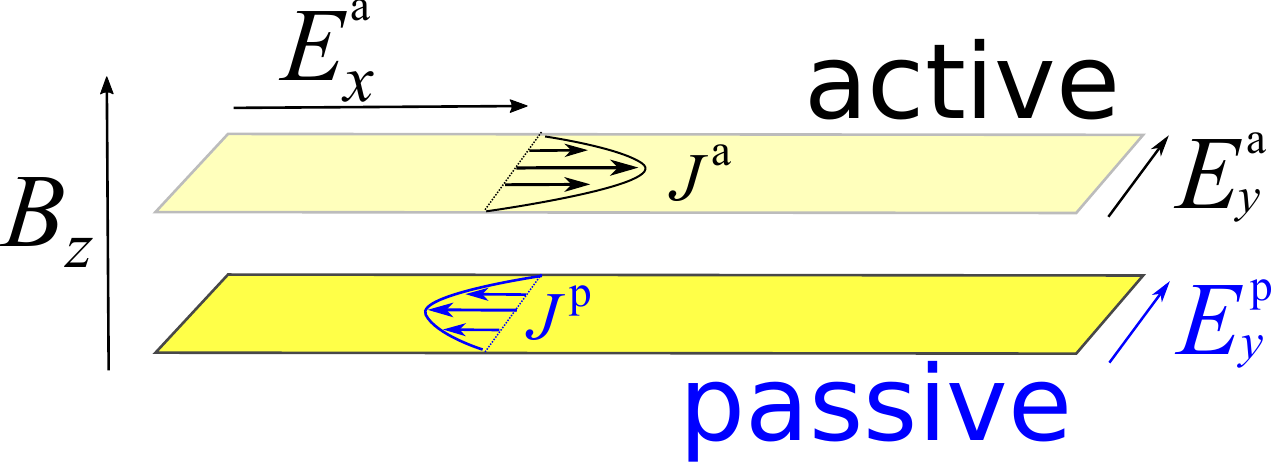}
\caption{\emph{Coulomb drag setup:} an electric field $E_x^\one \hat{\vec x}$ is applied to the active layer causing a Poiseuille current profile $J^\one (y) \hat{\vec x}$. This current induces electron motion in the passive layer controlled by the drag coefficient $\gamma_d^\two$ and the drag viscosity $\nu_d^\two$. When the electron density of the passive layer is much higher than that of the active layer, $\gamma_d^\two$ becomes very small and $J^\two$ changes sign. The magnetic field $B_z \hat{\vec z}$ and horizontal flow $J^\lambda (y) \hat{\vec x}$ cause a charge build-up and a transversal electric field $E_y^\lambda\hat{\vec y}$ that can be used to probe the drag-Hall viscosity $\nu_{dH}^\lambda$. \label{fig1} }
\end{figure}

In this work, we show that two new viscosity terms emerge indeed in the Coulomb drag magneto-transport of viscous fluids. For this purpose, we solve the coupled kinetic equations for the electrons in the two layers that interact via Coulomb interactions. The angular harmonics of the nonequilibrium distribution function give access to macroscopic quantities including the stress tensor in the linear-response, low-temperature limit (with Fermi energy $E_F\gg T$). The effects of intra- and interlayer Coulomb interactions on the stress tensor lead to the conventional viscosity and the drag viscosity, respectively, in the Navier-Stokes equations. In the presence of a magnetic field, those intra- and interlayer interactions will additionally induce the Hall and the drag-Hall viscosities.

We show that the resulting four viscosities are tunable by varying the ratio of the electron densities in the two layers and the magnetic field strength. Equipped with these four viscosities, we apply the Navier-Stokes equation to Coulomb drag in a Hall bar geometry, where the boundary conditions lead to Poiseuille flow. Such flow has been visualized in many experiments including graphene and Weyl semimetals~\cite{Sulpizio2019-poiseuille,Vool2021}. We show that at certain density ratios, the drag viscosity balances the stress force from the kinematic viscosity in the passive layer and becomes stronger than the drag force. This situation causes the electrons in the passive layer to flow opposite to the flow in the active layer, a phenomenon which gives rise to a negative drag conductivity (see Fig.~\ref{fig1}). Under an applied magnetic field, the transverse electric field shows a sign change due to the drag-Hall viscosity and is tunable by varying the density ratio.

\begin{figure*}[t]
\resizebox{2.0\columnwidth}{!}{
\includegraphics[height=3.5cm]{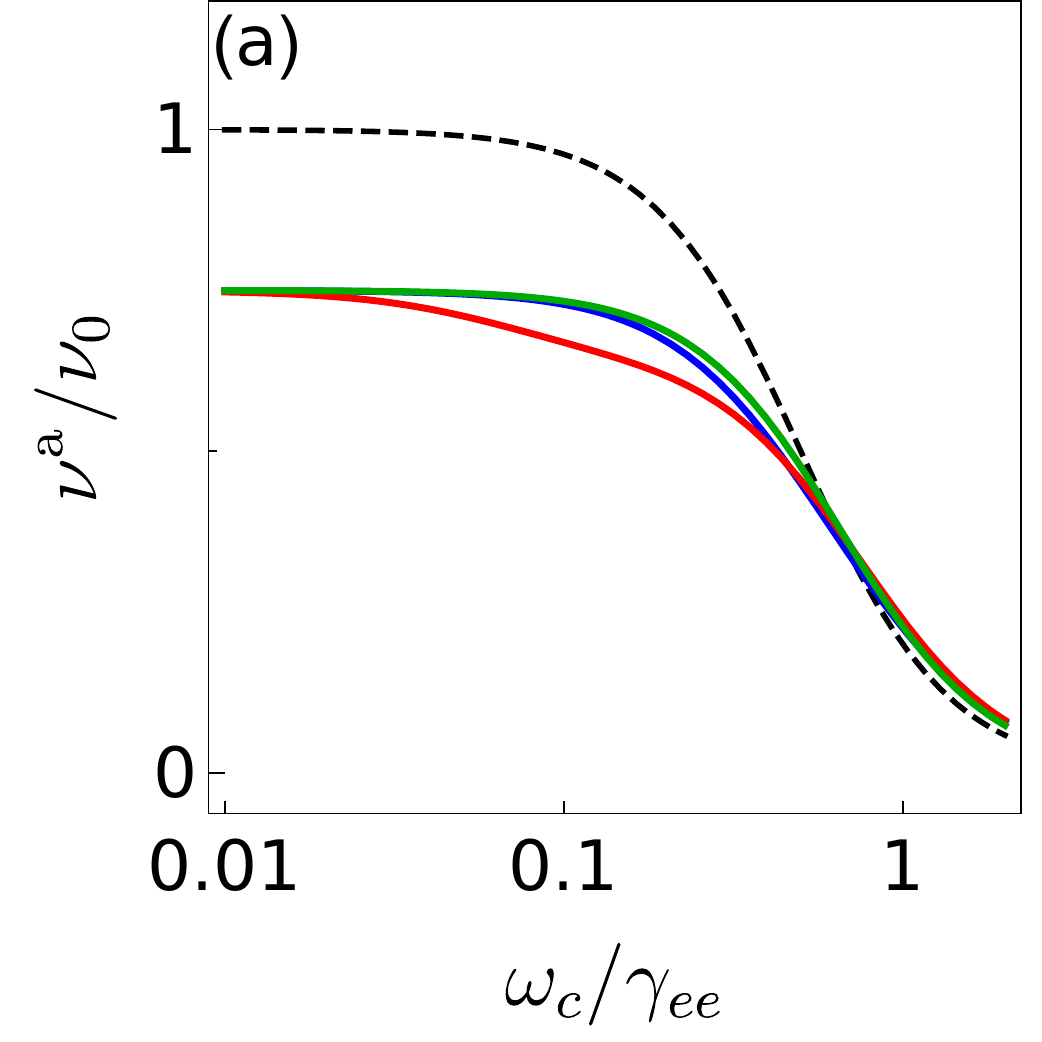}
\includegraphics[height=3.5cm]{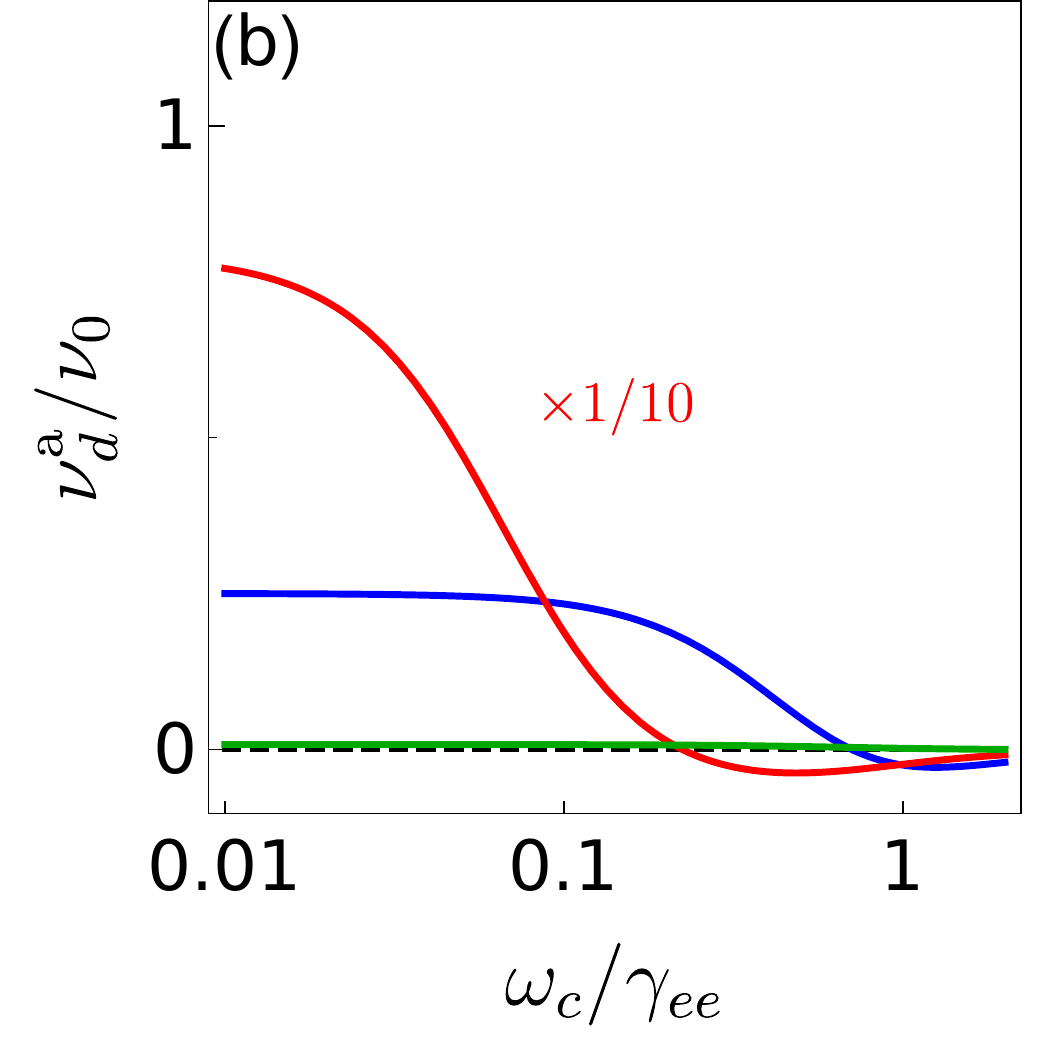}
\includegraphics[height=3.5cm]{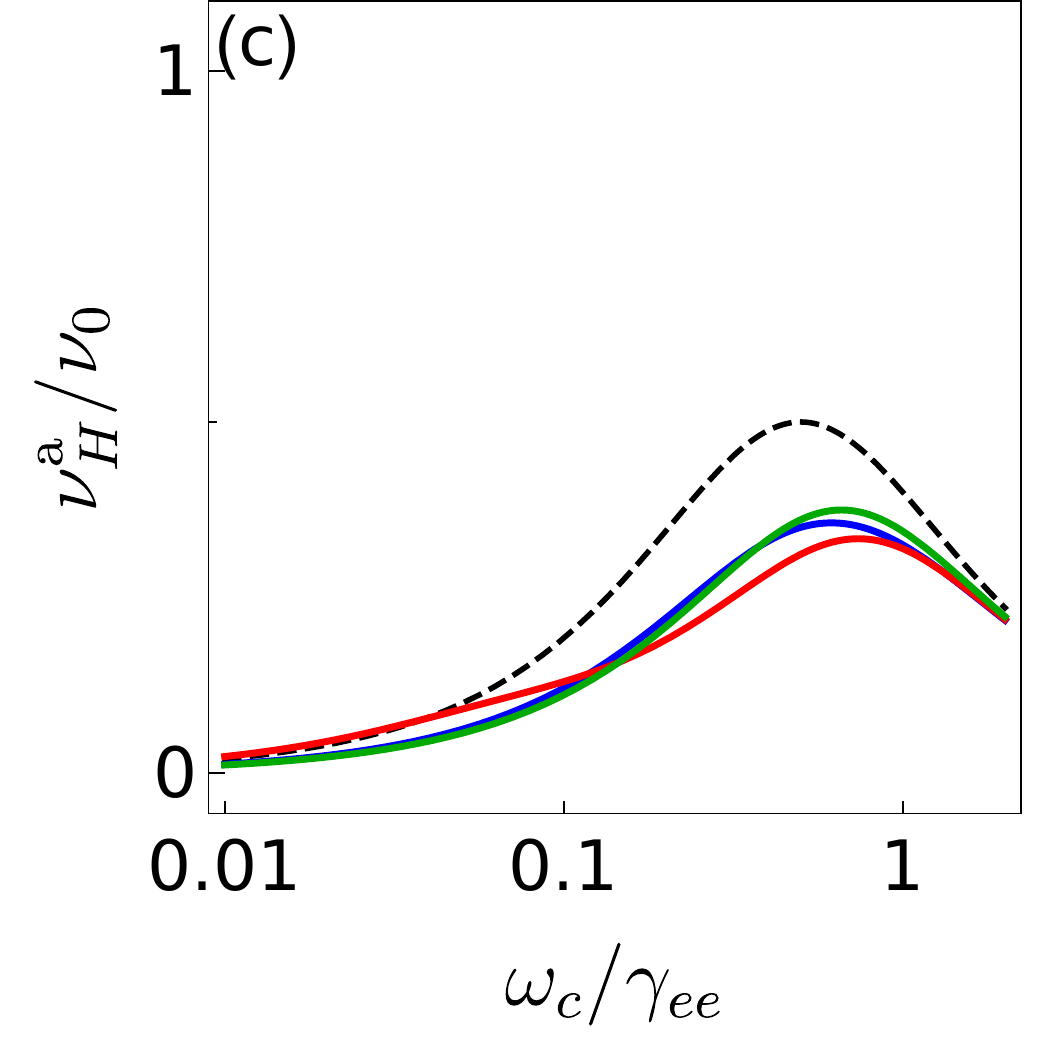}
\includegraphics[height=3.5cm]{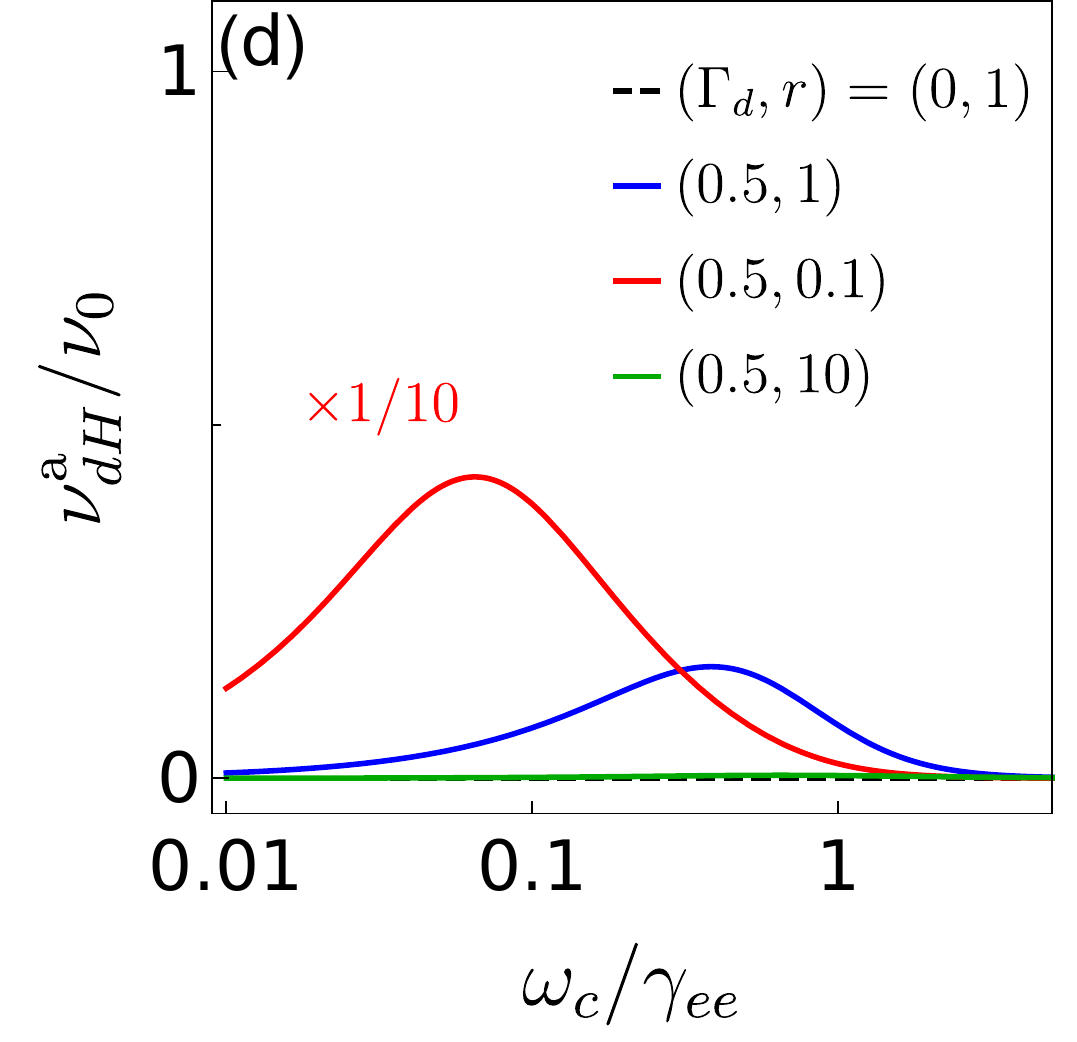}}
\resizebox{2.0\columnwidth}{!}{
\includegraphics[height=3.5cm]{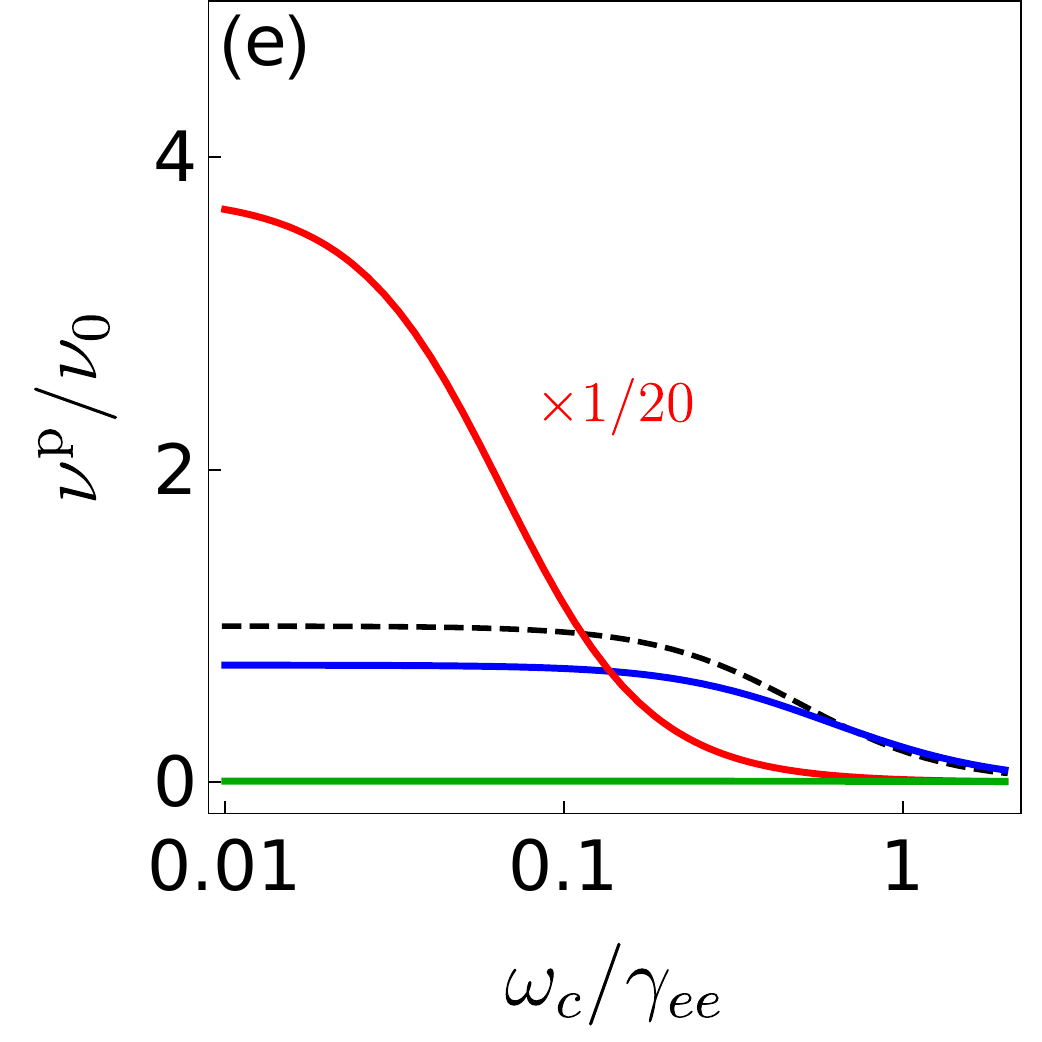}
\includegraphics[height=3.5cm]{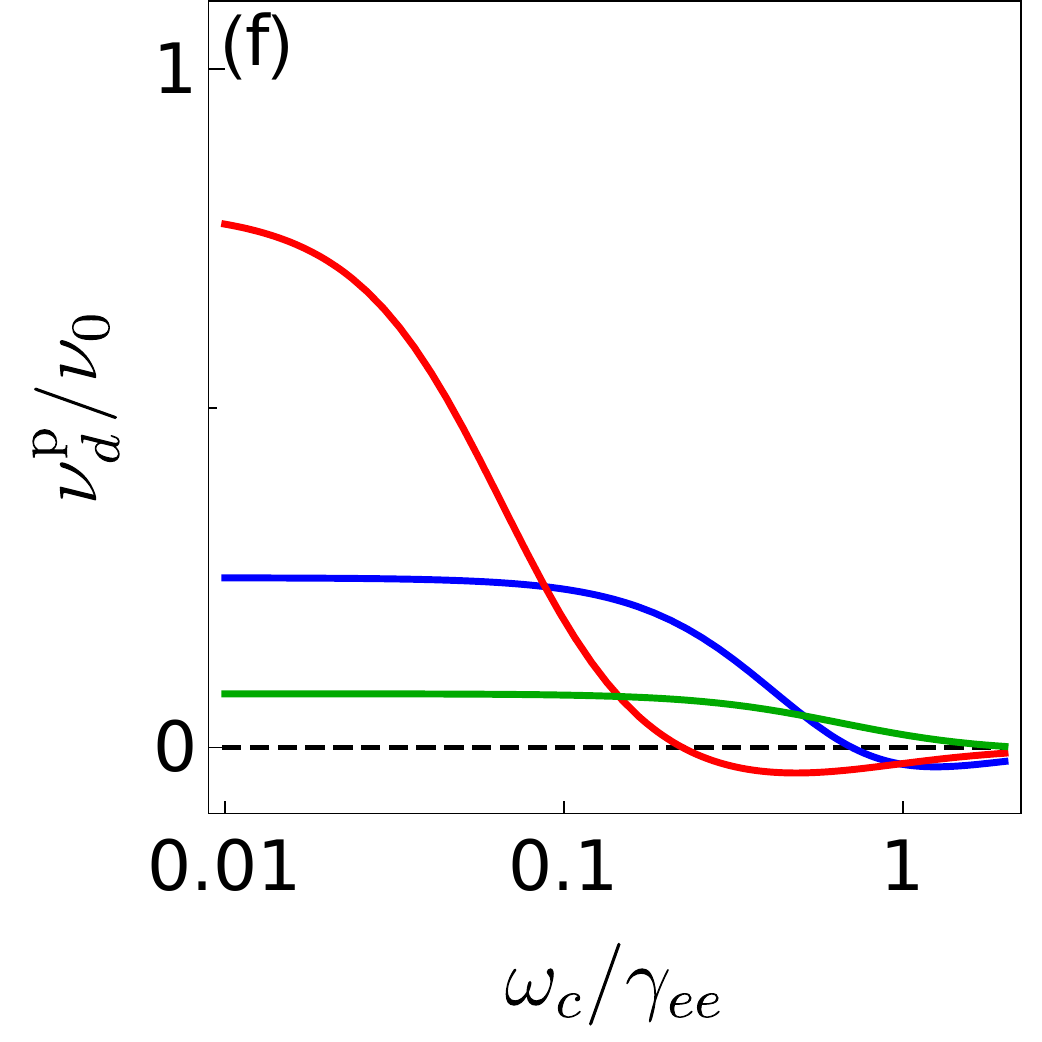}
\includegraphics[height=3.5cm]{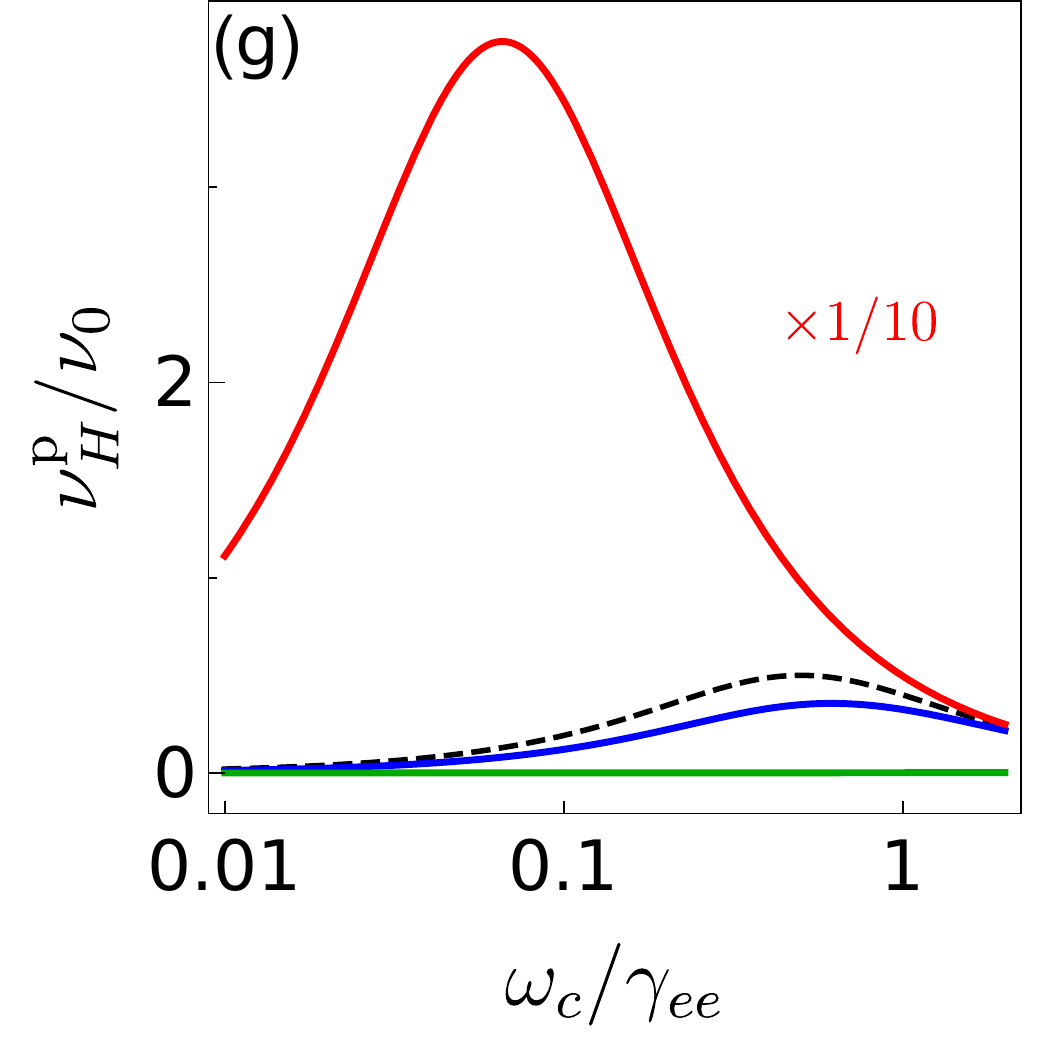}
\includegraphics[trim=1cm 0cm 0.5cm 0cm,clip,height=3.5cm]{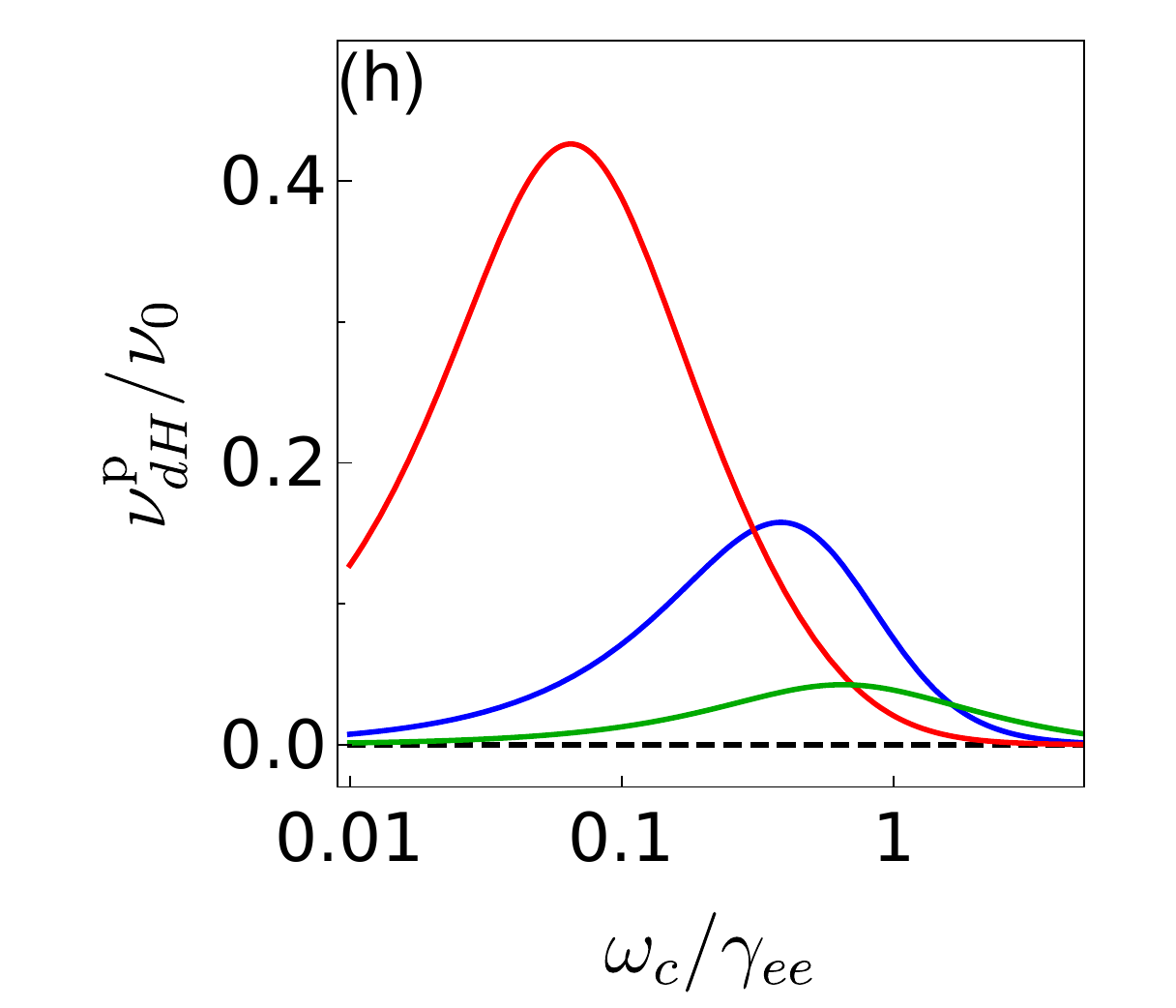}}
\caption{\label{fig:visco} The viscosities in both layers. (a,e) The kinematic viscosity $\nu^\lambda$, (b,f) the drag viscosity $\nu_d^\lambda$, (c,g) the Hall viscosity $\nu_H^\lambda$ and (d,h) the drag-Hall viscosity $\nu_{dH}^\lambda$ in the active and passive layers, respectively, for different magnetic fields, drag coefficients $\Gamma_d=\gamma_d/\gamma_{ee}$, and density ratios $r=\bar n_\one/\bar n_\two$. In Figs.(b), (c), (e), and (g), we have multiplied the red lines ($\Gamma_d=0.5$ and $r=0.1$) by the factors written in the plot. Here $\nu_0=(v_F^\one)^2/(4 \gamma_{ee}^\one)$. }
\end{figure*}

\emph{Viscosities due to Coulomb drag} ---
The system we consider consists of a pair of two-dimensional metallic layers separated by a distance much shorter than the screening length. To ensure hydrodynamic electron flow, we assume that the metals are clean such that the intra- and interlayer Coulomb scattering rates are much faster than those due to impurity and phonon scattering. Moreover, we consider an in-plane applied electric field in the active layer and allow for an applied out-of-plane magnetic field $\vec B =B_z \hat{\vec z}$ (see. Fig.~\ref{fig1}). The macroscopic dynamics of hydrodynamic electrons at small flow velocity can then be described by the linearized Navier-Stokes equation (NSE), derived in the Appendix,
\begin{align}
 \partial_t\vec u^\lambda&= \frac{e}{\alpha m} \boldsymbol{\mathcal E}^\lambda + \omega_c \vec u^\lambda \times \hat{\vec z} - \gamma_d^\lambda (\vec u^\lambda-\vec u^{\bar \lambda}) + f_{\rm visc}^\lambda, \label{eq:nse} \\
f^\lambda_{\rm visc} &= \nu^\lambda \nabla^2 \vec u^\lambda + \nu_H^\lambda \nabla^2\lp \vec u^\lambda \times \hat{\vec z} \rp \nn
&+\nu_d^\lambda \nabla^2 \vec u^{\bar \lambda} +\nu_{dH}^\lambda \nabla^2\lp \vec u^{\bar \lambda} \times \hat{\vec z} \rp,
\end{align}
where $\vec u^\lambda(\vec r,t)$ is drift velocity of electron in layer $\lambda \in \{\rm a,\ p\}$  (where we use $\bar{\lambda}$ to designate the opposite layer), $\gamma_d^\lambda$ is the rate of inter-layer scattering and known as the drag coefficient and  $e$ and $m$ are electron charge and band mass, respectively. As is shown in the Appendix, the equation is also valid for Dirac electrons, in which case $m = p_F/v_F$ corresponds to the effective cyclotron mass. The band parameter $\alpha=1$ for a parabolic band and $\alpha=2$ for a linear band. Moreover, $\nu^\lambda$ is the kinematic viscosity which is inversely proportional to the rate of intra-layer scattering $\gamma_{ee}^\lambda$, $\nu_H^\lambda$ is the Hall viscosity which is proportional to the cyclotron frequency $\omega_c=eB_z/mc$, $c$ being the speed of light. $\nu_d^\lambda\propto \gamma_d^\lambda$ is the drag viscosity, and $\nu_{dH}^\lambda \propto \omega_c\gamma_d^\lambda$ is the drag-Hall viscosity. The total electric field in a given layer is denoted by $\boldsymbol{\mathcal E}^\lambda= \vec E^\lambda + \frac{\alpha m}{e} \nabla P^\lambda=-\nabla \varphi^\lambda$, where $\vec{E}^\lambda$ is the externally applied electric field and $\nabla P^\lambda$ is the gradient of pressure. The presence of the drag viscosity $\nu_d^\lambda$ and the drag-Hall viscosity $\nu_{dH}^\lambda$ in the NSE is one of main results in this work.

The electron-electron interaction rate in a 2D electron gas is related to the density as $\gamma_{ee}^\lambda\propto T^2/E_F^\lambda\propto 1/\bar n_\lambda$, where  $\bar n_\lambda$ is the carrier density. The drag coefficient $\gamma_d^\lambda$ are tunable by varying interlayer spacing and the density of the opposite layer~\cite{gantmakher77-drag,abanin11-spin,justin13-drag,steve-simon20-qboltz-blg}. Furthermore, as derived in the Appendix, all four viscosities $\nu^\lambda,\ \nu_d^\lambda,\ \nu_H^\lambda,\ \nu_{dH}^\lambda$ are adjustable by varying the density ratio $r=\bar n_\one/\bar n_\two$, the ratio $\Gamma_d=\gamma_{d}^\one/\gamma_{ee}^\one$, and the strength of the magnetic field. Hereafter, we simplify the notation by scaling the quantities with the active layer such that $\gamma_{ee,d}^\one=\gamma_{ee,d}$, $\gamma_{ee,d}^\two=r \gamma_{ee,d}$, $\nu^\one=\nu$, and other coefficients as derived in Eqs.~(\ref{eq:nu1})-(\ref{eq:nu2}) of the Appendix.

In Fig.~\ref{fig:visco}(a), we show the kinematic viscosity of the active layer as a function of the magnetic field. The dashed line refers to the limit of vanishing drag coefficient $\gamma_d=0$ whereas the solid lines with different colors correspond to different density ratios $r = \bar n_\one/\bar n_\two$ and fixed $\Gamma_d=0.5$. We normalize all viscosities with respect to the kinematic viscosity at vanishing magnetic field and drag coefficient $\nu_0 \equiv \nu^\one(B_z=0,\gamma_d = 0) = v_F^2/(4\gamma_{ee})$~\cite{pellegrino17-hallvisc,ledwith19-visc}, where $v_F$ is the Fermi velocity of the active layer. In the presence of drag, the viscosity decreases similar to the effect of momentum relaxing scattering. The change of density in the passive layer does not change significantly the viscosity in the active layer as shown by the behavior of  $\nu^\one$ vs $r$.  At large magnetic fields, the viscosity decreases as shown in Ref.~\cite{alekseev16-hallvisc}, leading to a negative magnetoresistance in the viscous fluid. The viscosity of the passive layer in Fig.~\ref{fig:visco}(e) shows a similar $\omega_c$ dependence as $\nu^\one$ but it is proportional to $r^{-2}$ implying the shown density dependence of the viscosity. The drag viscosities $\nu_d^\lambda$ in Figs.~\ref{fig:visco}(b,f) vanish at zero drag $\gamma_d=0$ and strongly depend on density ratio $r$ with different dependencies in active and passive layers. $\nu_d^\lambda$ can even become negative at large enough magnetic fields $\omega_c\approx\gamma_{ee}$.

The Hall viscosity $\nu_H^\one$ shows a monotonic increase with $\omega_c$ as long as $\omega_c<\gamma_{ee}$.  At very large $\omega_c$, its value reduces. At small $\omega_c$, we can approximate $\nu_H^\lambda \propto \omega_c$. $\nu_H^\two$ shows a similar qualitative $\omega_c$ dependence as $\nu_H^\one$ with quantitative differences in the $r$ dependence due to the scaling with respect to the viscosity $\nu_0$ of the active layer [Fig.~\ref{fig:visco}(g)]. A nonzero drag Hall viscosity $\nu_{dH}^\lambda$ requires both $\omega_c$ and $\gamma_d$ to be simultaneously nonzero [Fig.~\ref{fig:visco}(d,h)]. By changing the density ratio $r$, the drag-Hall viscosity $\nu_{dH}^\one$ can be made larger than $\nu_H^\one$.

\emph{Effects on Poiseuille flow} ---
Next, we specialize Eq.~\eqref{eq:nse} to the case of steady-state Poiseuille flow in a narrow strip along the $x$-direction. We apply an electric field $\boldsymbol {\ft E}^\one=E_x^\one \hat{\vec x}$  in the active layer and set $\boldsymbol {\ft E}^\two=0$. In the presence of an applied vertical magnetic field, a transversal electric field $E_y^\lambda$ builds up that ensures zero Hall current ($u_y^\lambda=0$) at equilibrium as imposed by the boundary conditions. We obtain the following equations for the longitudinal component (see Appendix for details),
\bea
\gamma_d ( u^\one- u^\two)&=&\nu \partial_y^2  u^\one + \frac{\nu_d}{r\sqrt{r}} \partial_y^2  u^\two + \frac{e }{\alpha m} E^\one_x,\label{eq1}\\
r \gamma_d ( u^\two- u^\one)&=& \frac{\nu}{r^2}  \partial_y^2  u^\two +  \frac{\nu_d}{\sqrt{r}} \partial_y^2  u^\one,\label{eq2}
\eea
where we have made the ansatz that $\vec u^\lambda = (u^\lambda,0)$ and assumed a fully developed flow where $u^\lambda(y)$ is independent of $x$. In that case, $\nabla \to \partial_y$ and we have taken $\omega_c/\gamma_{ee}\ll 1$ to simplify the $r$ dependence of the coefficients. It is important to note that at small $\omega_c$, the effect of magnetic field is negligible in the longitudinal motion. For Dirac fermions with linear spectrum, one needs to replace $\sqrt{r} \to 1$, which is related to $v_F^\one/v_F^\two$, in the denominators of Eqs.~\eqref{eq1} and \eqref{eq2}. Hereafter, we focus only on the case of parabolic band. 

Examining the dynamics in the passive layer using Eq.~\eqref{eq2}, one finds that $u^\two$ will be parallel to $u^\one$ due to the drag force ($\propto \gamma_d$) if one disregards the $\nu_d$ term. In the presence of $\nu_d$ and at small $r$, however, the drag force becomes negligible in comparison to the viscosity term. As a result, it emerges from Eq.~(\ref{eq2}) that the $\nu_d$ term will enforce a balance of stress forces with the $\nu$ term, resulting in opposite curvatures of the velocity profiles $u^\one$ and $u^\two$ along the transversal direction $y$ [see Fig.~\ref{fig1}]. In the case of no-slip boundary condition at the edges, the velocities vanish at the edges such that $u^\lambda(\pm w_h)=0$, where $w_h$ is the half-width of the system, while the velocity reaches a maximum at the center, thus creating a parabolic Poiseuille profile along $y$. The opposite curvatures of the velocity profiles between the two layers entail that $u^\one(y)$ and $u^\two(y)$ have opposite signs. One might argue that the balance of stresses arising from $\nu$ and $\nu_d$ can be diminished by inducing a pressure gradient $\nabla P^\two$ or an internal electric field in the passive layer. However, this effect is weak in the limit where the flow is incompressible fluid and kept at a constant temperature along the flow.

For general values of $r$, the solutions of Eqs.~\eqref{eq1} and~\eqref{eq2} are $u^{\one,\two}=u_0 ({\tilde u}_+\pm {\tilde u}_-)/2$, where $u_0=eE_x^\one w_h^2/(m\nu)$  and
\bea
{\tilde u}_- &=&  \frac{1+ r^{3/2}\tilde \nu_d}{\tilde \gamma_d (1+r^3+2r^{3/2}\tilde\nu_d)} \lp 1- \frac{\cosh(q \tilde y)}{\cosh(q)}\rp, \label{eq:umin}\\
{\tilde u}_+ &=&   (1-\tilde y^2) \xi +\lp \frac{1-r^3}{1+r^3+2r^{3/2}\tilde\nu_d}\rp{\tilde u}_-,\label{eq:uplus}
\eea
and
\bea
\tilde q&=& \lp \frac{\tilde\gamma_d(1+r^3+2r^{3/2}\tilde\nu_d)}{(1-\tilde \nu_d^2)}\rp^{1/2}, \nn
\xi&=&\frac{r^3}{1+r^{3}+2r^{3/2}\tilde\nu_d}.
\eea
Here, we used the dimensionless parameters $\tilde y=y/w_h$, $\tilde \nu_d=\nu_d/\nu$, and $\tilde \gamma_d = \gamma_d w_h^2/\nu$ which are related to the Reynolds number ${\cal R}=\gamma_{ee} w_h^2/\nu$. Writing the averaged charge current across the transversal direction $y\in [-w_h,w_h]$ as $\la J^\lambda \ra = \bar n_\lambda e\la u^\lambda\ra/2$, where $\la\mathcal{O}\ra=\frac{1}{2}\int_{-1}^{1}d\tilde y \mathcal{O}$
and defining the normal and the drag conductivities as $\sigma^\lambda = \la J^\lambda \ra/E_x^\one$, we obtain
\bea
\sigma^\one&=&\frac{\bar n_\one e^2 w_h^2}{m\nu} \lb \frac{1}{3}\xi +\langle {\tilde u}_-\rangle\lp\frac{1+r^{3/2}\tilde \nu_d}{1+r^3+2r^{3/2}\tilde \nu_d}\rp\rb,\label{eq:sigma1}\\
\sigma^\two&=&\frac{\bar n_\two e^2 w_h^2}{m\nu} \lb \frac{1}{3}\xi -\langle {\tilde u}_-\rangle\lp\frac{r^3+r^{3/2}\tilde\nu_d}{1+r^3+2r^{3/2}\tilde\nu_d}\rp\rb.\label{eq:sigma2}
\eea

\begin{figure}[t]
 \resizebox{\columnwidth}{!}{\includegraphics [height=6cm]{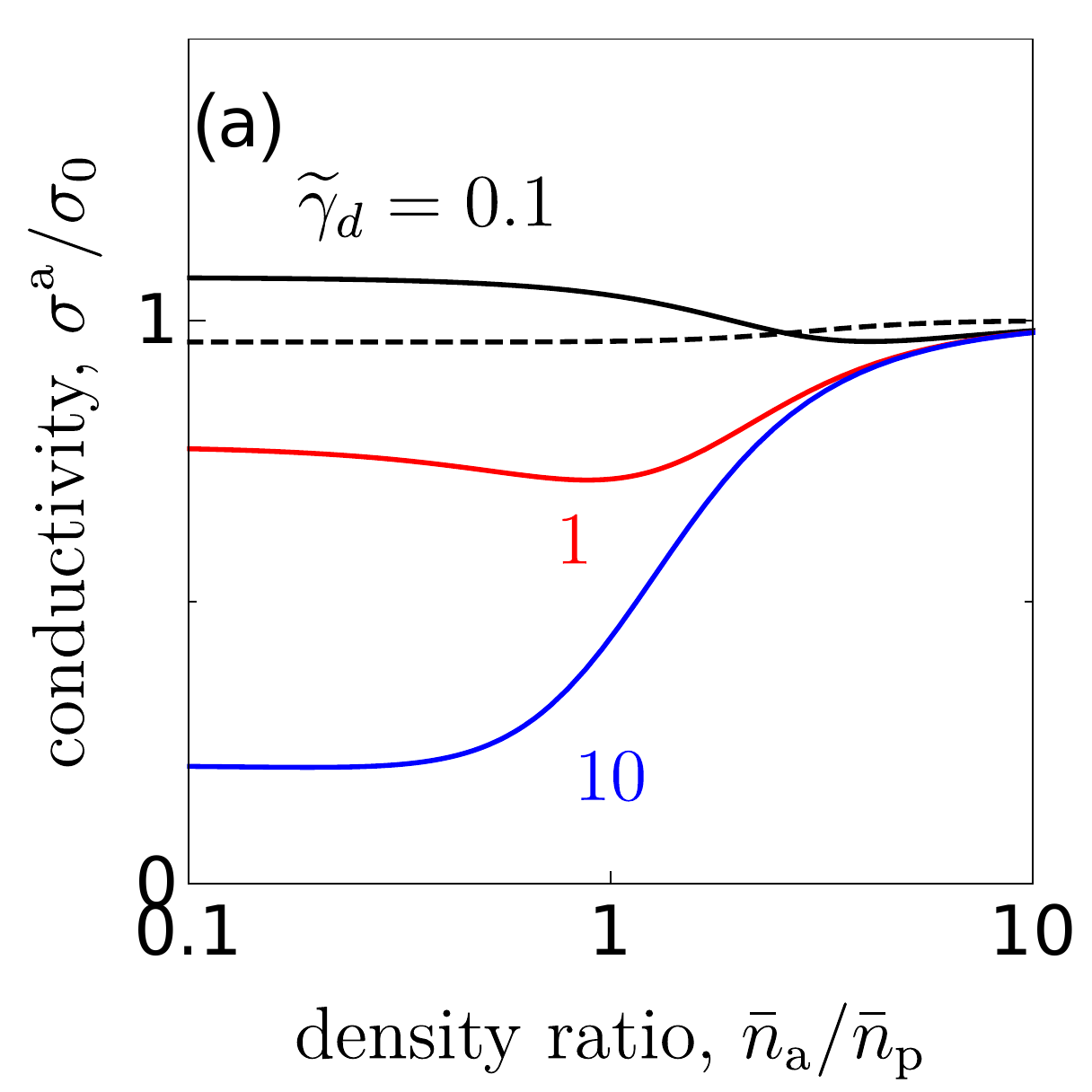}
 \includegraphics[trim=0cm 0.5cm 0cm  0cm, clip ,height=6cm] {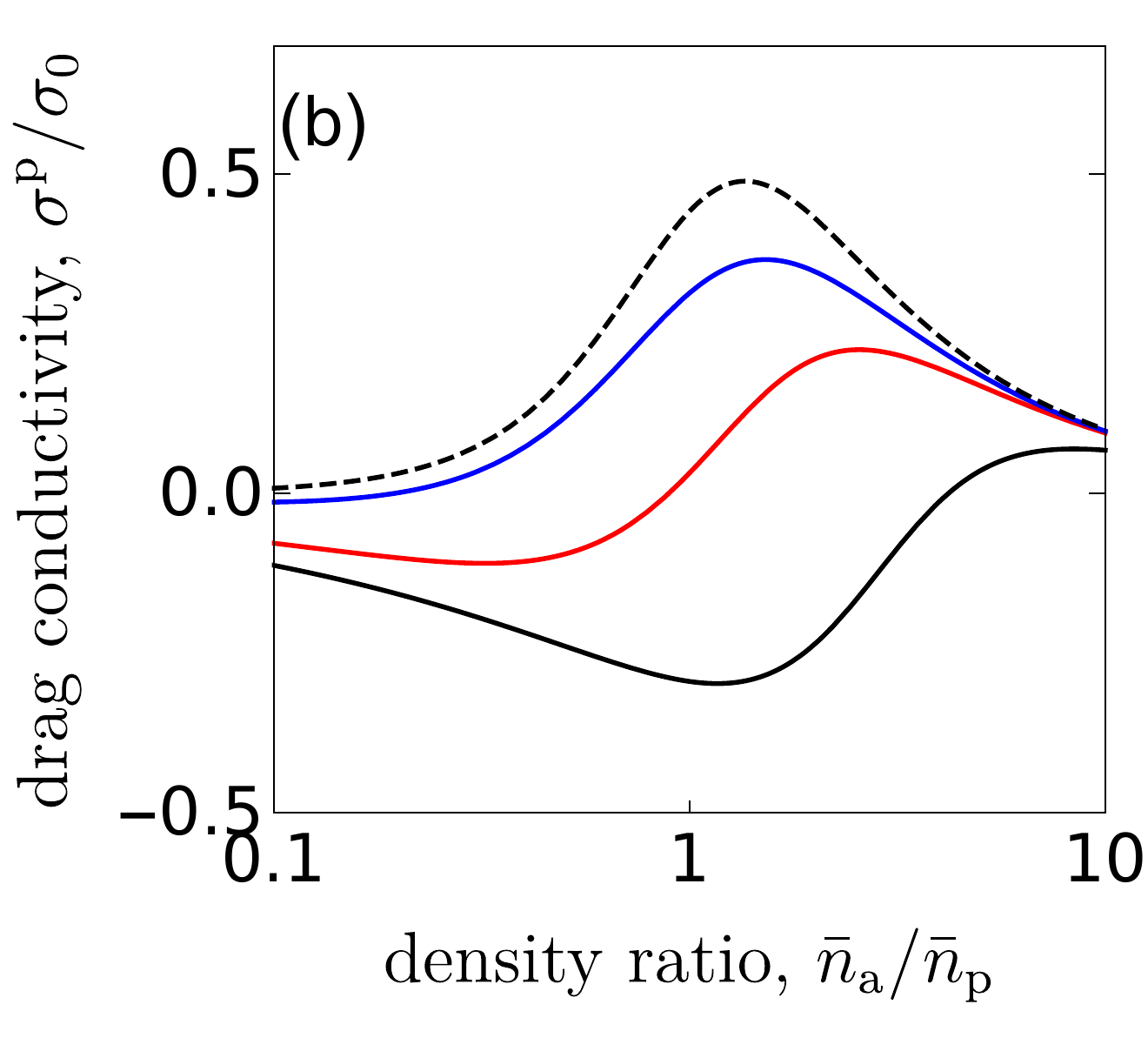}}
\caption{\label{fig:cond} (a) Conductivity $\sigma^\one$ and (b) drag conductivity $\sigma^\two$ as a function of $r=\bar n_\one/\bar n_\two$ for several values of $\tilde\gamma_d$. In this plot we have used $\nu_d/\nu=1/3$ and $\sigma_0=n_\one e^2w_h^2/(3 m \nu)$. The dashed lines correspond to $\tilde\gamma_d=0.1$ and neglect the drag viscosity. }
\end{figure}

In Fig.~\ref{fig:cond}, we plot $\sigma^{\one,\two}$ as a function of density ratio $r$ for several values of $\tilde\gamma_d$ and a fixed parameter $\tilde\nu_d=1/3$ corresponding to $\gamma_d/\gamma_{ee}=0.5$. The dashed lines correspond to the case when we neglect $\nu_d$.  By increasing $\tilde\gamma_d$, $\sigma^\one$ decreases indicating  the increase of the drag resistance. At small $r$, the drag resistance from the other layer is very strong leading to small values of $\sigma^\one$. In the non-viscous regime $\tilde \gamma_d \gg 1$ (the blue line), we can see a monotonic increase of $\sigma^\one$ as function of $r$ which saturates at $\sigma_0=n_\one e^2 w_h^2/(3 m\nu)$ for large $r$ where the effect of the drag force is minimal. The scale factor $\sigma_0$ is the conductivity of the viscous fluid without the drag effect, which takes the shape of a Drude conductivity where the effective lifetime depends on the channel width and the viscosity $\tau_\nu=w_h^2/(3\nu)$. In the highly viscous regime, $\tilde\gamma_d\ll 1$ (black line), the drag viscosity can enhance the conductivity at small $r$ originating from the second term of Eq.~\eqref{eq:sigma1} [cf.~the dashed line when $\nu_d=0$].


The impact of $\nu_d$ is most pronounced for the drag conductivity $\sigma^\two$, see Fig.~\ref{fig:cond}(b). At small density ratio $r$ and in the viscous regime $\tilde \gamma_d<1$, $\sigma^\two$ becomes negative signifying a counterflow in the passive layer with respect to the active one. At $r=0$, $\sigma^\two$ becomes zero because the drag coefficient $r\gamma_d$ vanishes, and it vanishes as well as at very large $r$ because $n_\two \to 0$, see Eq.~\eqref{eq:sigma2}. A negative $\sigma^\two$ occurs in the viscous regime when the effect of $\gamma_d$ term is smaller than those of the $\nu$ and $\nu_d$ terms, see Eq.~\eqref{eq2}, causing opposite signs of $u^\one$ and $u^\two$ due to the stress balance. Indeed, when we set $\nu_d=0$, $\sigma^\two$ never reaches a negative value (dashed line). Overall, the value of $|\sigma^\two|$ is typically smaller than that of $\sigma^\one$ but at large $\tilde \gamma_d$, $\sigma^\two/\sigma^\one$ can approach unity at $r=1$.

\begin{figure}[t]
\resizebox{\columnwidth}{!}{\includegraphics{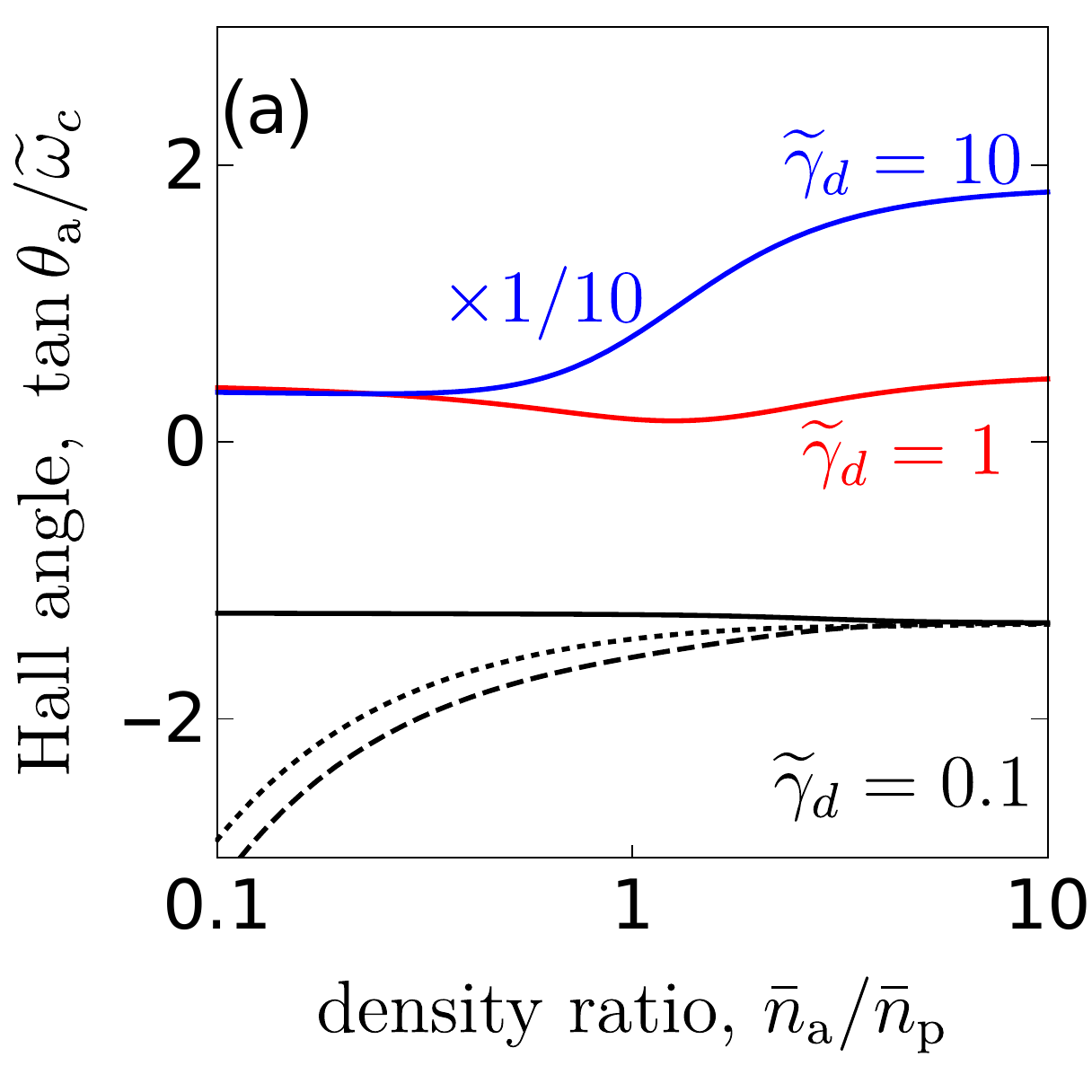},\hspace{5mm},\includegraphics{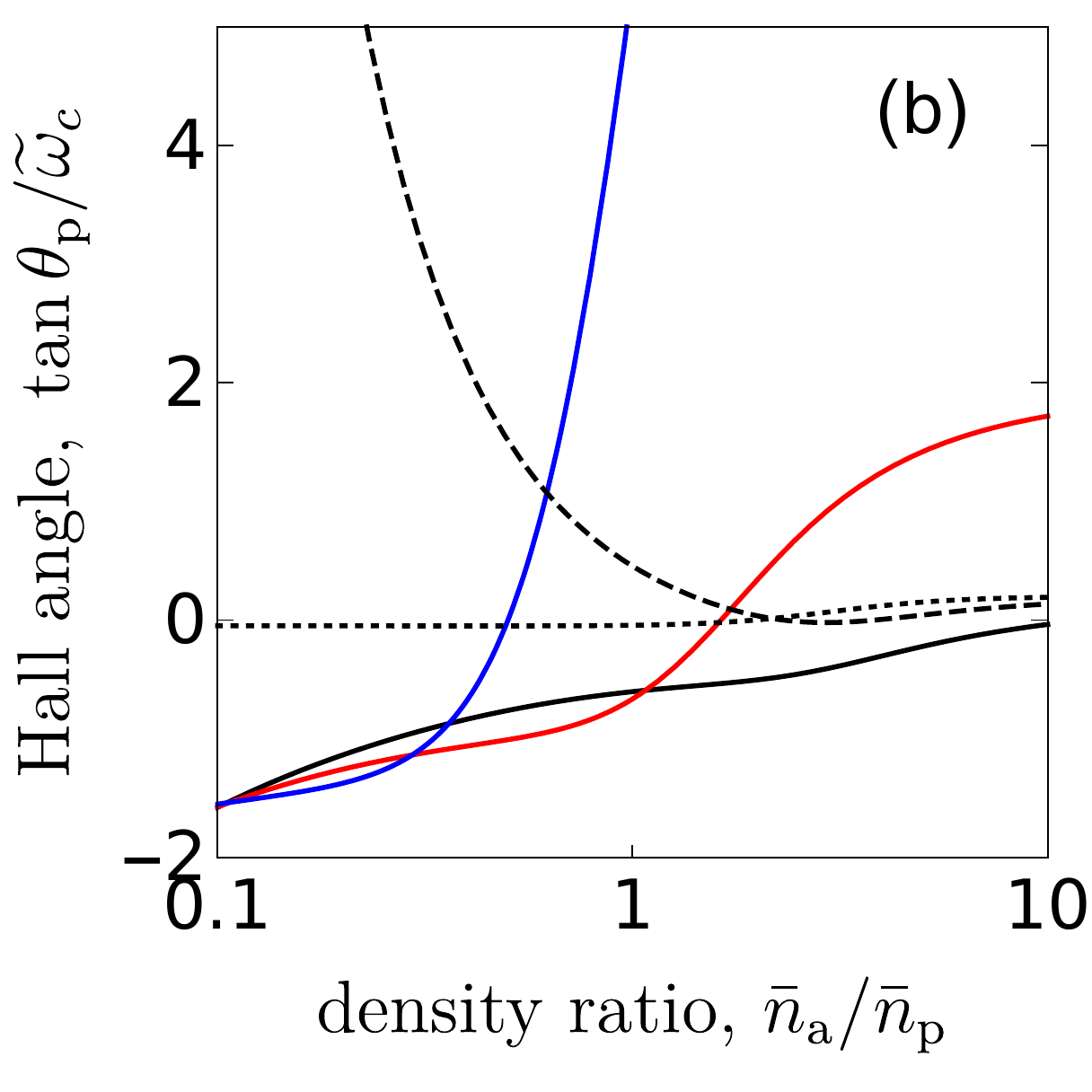}}
\caption{\label{fig:Hall angle} (a) Hall angles at the first layer $\tan\theta_\one/\tilde{\omega}_c$ and  (b) at the the second layer $\tan\theta_\two/\tilde{\omega}_c$  as a function of $r=\bar n_\one/\bar n_\two$ for several values of $\tilde\gamma_d=\Gamma_d\mathcal{R}$. Here $\tilde\omega_c=\omega_c/\gamma_{ee}$ and $\Gamma_d=0.5$. The dashed lines are for $\tilde\gamma_d=0.1$ and neglecting the drag-Hall viscosity $\nu_{dH}=0$ and the dotted lines are for $\tilde\gamma_d=0.1$ and $\nu_{dH}=0$ and $\nu_d=0$. }
\end{figure}

One can also consider the transverse component of Eq.~\eqref{eq:nse} and relate the velocity along the strip $u^\lambda$ with the perpendicular electric field $E_y^\lambda$ due to the magnetic field,
\be
\nu_H^\lambda \partial_y^2 u^\lambda + \nu_{dH}^\lambda \partial_y^2 u^{\bar \lambda} =
\frac{e }{m} E_y^\lambda - \omega_c u^\lambda. \label{eq:transverse}
\ee
Since $u^\lambda$ is proportional to $E_x^\one$, one can measure the Hall angle $\tan\theta_\lambda=E_y^\lambda/E_x^\one$. Utilizing Eqs.~\eqref{eq:umin} and \eqref{eq:uplus}, we obtain $\tan\theta_\lambda$ as shown in Fig.~\ref{fig:Hall angle}. At small magnetic field $\omega_c/\gamma_{ee}=\tilde\omega_c\ll 1$, $\tan\theta_\lambda$ is linearly proportional to the strength of magnetic field  $\tilde \omega_c$. The polarity of $E_y^\one$ is less sensitive to $r$ but more sensitive with the change of the Reynolds number represented by $\tilde\gamma_d$. On the other hand in the passive layer, $E_y^\two$ changes the sign by changing $r$ but less sensitive to the change of $\tilde \gamma_d$. The presence or absence of $\nu_d$ and $\nu_{dH}$ also change the polarity of $E_y^\two$ as shown in the dashed and dotted lines in comparison to the solid line.

\emph{Conclusion} --- Starting from the kinetic Boltzmann equation for two metallic layers interacting via the Coulomb interaction, we have shown that viscous hydrodynamic transport in such a Coulomb drag setup is characterized by four viscosities: the kinematic, Hall, drag, and drag-Hall viscosities. Those viscosities are tunable by several parameters such as the applied magnetic field, the charge density ratio in the layers, and the drag coefficient (interlayer spacing). We showed that the drag viscosity can lead to a counterflow between electrons in the passive layer and the active one in the viscous regime. This phenomenon can be measured via a negative drag conductivity $\sigma^\two$ and is independent of the magnetic field in the small field regime $\omega_c\ll\gamma_{ee}$. In the presence of magnetic field, the polarity of the Hall fields $E_y^\lambda$ can be altered by varying the Reynolds number and the density ratio to probe the presence of the drag-Hall viscosity.

\emph{Acknowledgements} --- The authors acknowledge support from the National Research Fund Luxembourg under grants CORE C20/MS/14764976/TOPREL, CORE C21/MS/15752388/NavSQM, and CORE C19/MS/13579612/HYBMES.

\appendix

\begin{widetext}

\section{Derivation of four viscosities}
\label{sec:derivation}
In a Coulomb drag experiment, we consider electrons in two layers $\lambda$ and $\bar \lambda$. Both intra-layer and inter-layer Coulomb interactions are strong, such that electrons quickly relax into local equilibrium distributions. Subject to such interactions and applied external forces, the electron dynamics follows the Boltzmann equation
\be
 \partial_t f^\lambda+\vec v_\vec p^\lambda \cdot \partial_{\vec r} f^\lambda +\lp e \vec E^\lambda+ \frac{e}{c}\vec v_\vec p ^\lambda \times \vec B \rp \cdot \partial_\vec p f^\lambda = S_{\lambda,\lambda}^e  + S_{\lambda,\bar \lambda}^d,\label{eq:kinetic}
 \ee
 where
 $\vec v_\vec p^\lambda=d\eps_\vec p^\lambda/d\vec p$ is the group velocity of $\lambda\in\{ \one,\ \two\}$ electrons in the active or passive layer, respectively. $\vec E^\lambda$ and $\vec B=\hat{\vec z}B_z$ are the electric and magnetic fields, respectively. The electric field may differ between the two layers while the magnetic field is assumed to be identical in both layers. In this derivation, we incorporate only the two dominant types of collision, namely intra- and interlayer electron-electron collisions, corresponding respectively to the collision integrals $ S_{\lambda,\lambda}^e$ and $S_{\lambda,\bar \lambda}^d$. Other momentum-relaxing collisions can be incorporated easily. We assume that the intralayer Coulomb interaction $S_{\lambda,\lambda}^e$ is the strongest of the problem, such that we can make the following ansatz for the local equilibrium distribution function \cite{pellegrino17-hallvisc},
 \bea
 f^\lambda (\vec r,\vec p,t) &=& f_0^\lambda(\eps_\vec p ^\lambda) - \frac{\partial f_0^{\lambda}(\eps_\vec p^\lambda)}{\partial \eps_\vec p^\lambda} F^\lambda (\vec r, \theta_\vec p,t ),\label{eq:fdist}
 \eea
 where $f_0^\lambda=\lp 1+\exp[\beta(\eps_\vec p^\lambda-\mu)]\rp^{-1}$ is the equilibrium Fermi distribution. We assume an isotropic and sharp Fermi surface at low temperatures. Therefore, the factor $\partial f^\lambda_0/\partial \eps_\vec p^\lambda$ is strongly peaked at the Fermi momentum. Hence, using $\vec{p} = |\vec{p}| (\cos \theta_{\vec p}, \sin \theta_{\vec p})$, we can assume $|\vec p| \approx p_F$ in the correction, such that the nonequilibrium distribution $F(\vec r,\theta_\vec p,t)$ depends on $\vec p$ mostly via the polar angle $\theta_\vec p$. Therefore, we can expand it in angular harmonics,
 \be
 F^\lambda(\vec r,\theta_\vec p,t) =\sum_{n= -\infty}^\infty e^{i n \theta_\vec p} \ft F_n^\lambda(\vec r,t).
 \ee
 We see that $\ft F_0^\lambda$ is related to the density fluctuations
 \bea
 n_\lambda (\vec r,t) &=& \int d^2\vec p  \lp f^\lambda(\vec r,\vec p,t)-f_0^\lambda(\eps_\vec p^\lambda)\rp
 =  g_F \ft F_0^\lambda (\vec r,t),\label{eq:dens}
 \eea
 where $ g_F = \int d^2\vec{p} \delta (\eps_F - \eps_{\vec{p}}^\lambda)$ is the local density of states at the Fermi level.
 Moreover, the functions $\ft F_{\pm 1}^\lambda$ are related to the current density,
 \bea
 \vec J^\lambda (\vec r,t) &=& \int d^2 \vec p\ \vec v_\vec p^\lambda \lp f^\lambda(\vec r,t)-f_0^\lambda(\eps_\vec p^\lambda,t)  \rp,\nn
 &=& \frac{1}{2} g_F v_F^\lambda \begin {pmatrix}
   \ft F_1^\lambda (\vec r,t) + \ft F_{-1}^\lambda (\vec r,t)\\
   i \left[ \ft F_1^\lambda (\vec r,t) - \ft F_{-1}^\lambda (\vec r,t)\right]
 \end {pmatrix}\equiv\bar n_\lambda \vec u^\lambda(\vec r,t), \label{eq:current}
 \eea
 where $\bar n_\lambda = \int d^2\vec{p} f_0^\lambda (\epsilon_\vec{p})=g_F\eps_F^\lambda$ is the equilibrium density and this equation defines the drift velocity $\vec u^\lambda(\vec{r}, t)$.
 The functions $\ft F_{\pm 2}^\lambda$ are related to the stress tensor,
 \bea
 \bar { T}^\lambda_{i,j} &=& \int d^2 \vec p\ p_i  v_{p,j}^\lambda (f^\lambda(\vec r,t) -f^\lambda_0(\eps_\vec p^\lambda)),\nn
 \bar { T}^\lambda_{x,x} 
 &=& \frac{g_F p_Fv_F^\lambda}{4}\lp 2\ft F_0^\lambda (\vec r,t) + \ft F_2^\lambda(\vec r,t) + \ft F_{-2}^\lambda(\vec r,t)\rp,\nn
 \bar T^\lambda_{x,y} =\bar T^\lambda_{y,x}&=& \frac{g_F p_Fv_F^\lambda}{4i} \lp \ft F^\lambda_{-2}(\vec r,t) -\ft F_{2}^\lambda(\vec r,t)\rp,  \\
 \bar T^\lambda_{y,y} &=& \frac{g_F p_F v_F^\lambda}{4}\lp 2\ft F_0^\lambda (\vec r,t) - \ft F_2^\lambda(\vec r,t) - \ft F_{-2}^\lambda(\vec r,t)\rp,\nonumber\eea
 We can express this tensor in a compact form as
 \be
 \bar {\vec T}^\lambda = \frac{g_F p_F v_F^\lambda}{4} \lb 2\ft F_0^\lambda + (\ft F_2^\lambda +\ft F_{-2}^\lambda)\tau_z+ i(\ft F_2^\lambda-F_{-2}^\lambda)\tau_x\rb,\label{eq:stress}
 \ee
 where $\tau_{x,z}$ are the Pauli matrices.
 Next, we model $S_{\lambda,\lambda}^e$ so that it conserves the particle number and momentum
 \be
 S_{\lambda,\lambda}^e=-\gamma_{ee}^\lambda \lp  F^\lambda(\vec r,\theta_\vec p,t) -\ft F_0^\lambda-\ft F_1^\lambda-\ft F_{-1}^\lambda\rp,
 \ee
 where $\gamma_{ee}=\tau_{ee}^{-1}$ is the electron-electron scattering rate. Recent works show the dependence of $\gamma_{ee}$ on odd and even harmonics $n$ at the crossover between hydrodynamics and ballistic transport~\cite{ledwith19-visc}. In this work, we focus on the hydrodynamic regime and assume that $\gamma_{ee}$ is independent of $n$.
 On the other hand, the interlayer collision integral $S_{\lambda,\bar \lambda}^d$ will give rise to drag effects that only conserve the particle number but not necessarily the momentum in a given layer. The collision integral is derived in Eq.~\eqref{eq:sdlin} and takes the following form
 \be
 S_{\lambda,\bar \lambda}^d=-\gamma_d^\lambda (\vec u^\lambda -\vec u^{\bar \lambda})=-\gamma_d^\lambda \lp F^\lambda (\vec r,\theta_\vec p,t) - \ft F_0^\lambda -\frac{v_F^{\lambda}}{v_F^{\bar \lambda}}\lp F^{\bar \lambda} (\vec r, \theta_\vec p,t) -\ft F_0^{\bar \lambda}\rp \rp
 \ee
 Inserting Eq.~\eqref{eq:fdist} into Eq.~\eqref{eq:kinetic} and keeping only linear terms in $\vec E$ and $F^\lambda$, we obtain:
 \be
 \lp -\frac{\partial f_0^\lambda}{\partial \eps_\vec p^\lambda}\rp \bigg[\partial_t F^\lambda (\vec r,\theta_\vec p,t)+  \vec v_\vec p^\lambda \cdot \lp \partial_{\vec{r}} F^\lambda (\vec r,\theta_\vec p,t) -e\vec E\rp + \frac{eB_z}{c} \frac{\left|\vec v_p^\lambda\right|}{|\vec p|} \partial_{\theta_\vec p} F^\lambda(\vec r,\theta_\vec p,t)\bigg]= S_{\lambda,\lambda}^e+S_{\lambda,\bar \lambda}^d \label{eq:kinetic2}
 \ee
 Next, we multiply Eq.~\eqref{eq:kinetic2} on both sides by $e^{in \theta_\vec p} $ and integrate over $d^2\vec p$. Hence, we obtain:
 \bea
 \partial_t\ft F_n^\lambda+\frac{v_F^\lambda}{2} \lb \partial_x \lp \ft F_{n-1}^\lambda +\ft F_{n+1}^\lambda\rp -i \partial_y  \lp \ft F_{n-1}^\lambda -\ft F_{n+1}^\lambda\rp\rb &&\nn
 +(-e)\frac{v_F^\lambda}{2} \lp E_x (\delta_{n,1}+\delta_{n,-1})-iE_y (\delta_{n,1}-\delta_{n,-1})\rp -i n \omega_c \ft F_n^\lambda &=&-\gamma_{ee}^\lambda \lp \ft F_n^\lambda-\ft F_0^\lambda\delta_{n,0}-\ft F_1^\lambda\delta_{n,1}-\ft F_{-1}^\lambda \delta_{n,-1}\rp\nn
 &&-\gamma_{d}^\lambda \lb \ft F_n^\lambda-\ft F_0^\lambda\delta_{n,0}-\frac{v_F^\lambda}{v_F^{\bar \lambda}}\lp\ft F_n^{\bar \lambda}-\ft F_{0}^{\bar \lambda} \delta_{n,0}\rp\rb,\label{eq:recursive}
 \eea
 where $\omega_c=eB_z/mc$ is the cyclotron frequency and $m=p_F^\lambda/v_F^\lambda$. Eq.~\eqref{eq:recursive} is the main ingredient to obtain macroscopic dynamics of the system.
 Taking $n=0$ and using the definition~\eqref{eq:current}, we obtain the continuity equation,
 \be
 \partial_t n_\lambda(\vec r,t)+\nabla\cdot \vec J^\lambda (\vec r,t) =0.
 \ee
 For $n=\pm1$ we get, respectively,
\begin{align}
 \partial_t\ft F_1^\lambda+\frac{v_F^\lambda}{2} \lb \partial_x \lp \ft F_{0}^\lambda +\ft F_{2}^\lambda\rp -i \partial_y  \lp \ft F_{0}^\lambda -\ft F_{2}^\lambda\rp\rb -e\frac{v_F^\lambda}{2} \lp E_x -iE_y \rp -i  \omega_c \ft F_1^\lambda &=
 -\gamma_{d}^\lambda \lp \ft F_1^\lambda-\frac{v_F^\lambda}{v_F^{\bar \lambda}}\ft F_1^{\bar \lambda}\rp,\label{eq:n=1} \\
 \partial_t\ft F_{-1}^\lambda+\frac{v_F^\lambda}{2} \lb \partial_x \lp \ft F_{-2}^\lambda +\ft F_{0}^\lambda\rp -i \partial_y  \lp \ft F_{-2}^\lambda -\ft F_{0}^\lambda\rp\rb -e\frac{v_F^\lambda}{2} \lp E_x +iE_y \rp +i  \omega_c \ft F_{-1}^\lambda &=-\gamma_{d}^\lambda \lp \ft F_{-1}^\lambda-\frac{v_F^\lambda}{v_F^{\bar \lambda}}\ft F_{-1}^{\bar \lambda}\rp,\label{eq:n=-1}
 \end{align}
 Adding Eq.~\eqref{eq:n=1} and \eqref{eq:n=-1} and multiplying with $v_F^\lambda/(2 \eps_F^\lambda)$, we get
\bea
\partial_t (\ft F_1^\lambda+\ft F_{-1}^\lambda) +\frac{v_F^\lambda}{2} \lb \partial_x \lp 2 \ft F_{0}^\lambda +\ft F_{2}^\lambda  +\ft F_{-2}^\lambda\rp -i \partial_y  \lp \ft F_{-2}^\lambda -\ft F_{2}^\lambda\rp\rb&&\nn
-e v_F^\lambda  E_x  -i  \omega_c (\ft F_1^\lambda -\ft F_{-1}^\lambda)&=&
 -\gamma_{d}^\lambda \lb \ft F_1^\lambda+\ft F_{-1}^\lambda-\frac{v_F^\lambda}{v_F^{\bar \lambda}}\lp \ft F_1^{\bar \lambda}+\ft F_{-1}^{\bar \lambda}\rp\rb,\nn
 \partial_t u_x^\lambda +\frac{1}{m\bar n_\lambda} \lb \partial_x \bar T_{xx}^\lambda+ \partial_y \bar T_{xy}^\lambda \rb -\frac{e }{\alpha m}  E_x  -  \omega_c u_y^\lambda &=&
 -\gamma_{d}^\lambda \lp u_x^\lambda-u_x^{\bar \lambda}\rp,
 \eea
 where $\alpha=1$ for a quadratic dispersion $\eps_F^\lambda=m \lp v_F^\lambda\rp ^2/2$ and $\alpha =2$ for a linear dispersion $\eps_F^\lambda = p_F v_F^\lambda =m \lp v_F^\lambda\rp ^2 $. This $\alpha$ factor will not change any results since it only renormalizes $u_0$ above Eq.~(5) in the main text. Hereafter, we focus on the quadratic dispersion.

 Subtracting Eq.~\eqref{eq:n=-1} from \eqref{eq:n=1} and multiplying by $i v_F^\lambda/(2\eps_F^\lambda)$ we get,
 \be
 \partial_t u_y^\lambda +\frac{1}{m\bar n_\lambda} \lb \partial_x \bar T_{yx}^\lambda+ \partial_y \bar T_{yy}^\lambda \rb -\frac{e }{m}  E_y  +  \omega_c u_x^\lambda =
 -\gamma_{d}^\lambda \lp u_y^\lambda -u_y^{\bar \lambda}\rp
 \ee
 Using $n=\pm 1$ in Eq.~\eqref{eq:recursive} and using Eq.~\eqref{eq:current} and~\eqref{eq:stress}, we find the linearized Navier-Stokes equation

 \be
  \partial_t\vec u^\lambda +\frac{1}{m \bar n_\lambda}\nabla \cdot \bar{\vec T}^\lambda  -\frac{e  }{m} \vec E-  \omega_c \lp  \vec u^\lambda \times \hat {\vec z}\rp = - \gamma_d^\lambda\lp \vec u^\lambda-\vec u^{\bar \lambda} \rp.
 \ee
We can approximately close the recursion equation~\eqref{eq:recursive} by setting $\ft F_n=0$ for $|n|\ge 3$~\cite{pellegrino17-hallvisc,Lucas2018}. For $n=2$ and focusing on the stationary distribution with $\partial_t \ft F_n=0$, we obtain
\bea
\frac{v_F^\lambda}{2} \lp \partial_x -i\partial_y \rp  \ft F_{1}^\lambda  -i 2 \omega_c \ft F_2^\lambda &=&-\gamma_{ee}^\lambda  \ft F_2^\lambda-\gamma_{d}^\lambda \lp \ft F_2^\lambda-\frac{v_F^\lambda}{v_F^{\bar \lambda}}\ft F_2^{\bar \lambda}\rp,\nn
 \frac{v_F^\lambda}{2} \lp \partial_x -i\partial_y \rp  \ft F_{1}^{\bar \lambda}  -i 2 \omega_c \ft F_2^{\bar \lambda} &=&-\gamma_{ee}^{\bar \lambda}  \ft F_2^{\bar \lambda}-\gamma_{d}^{\bar \lambda} \lp \ft F_2^{\bar \lambda}-\frac{v_F^{\bar \lambda}}{v_F^{\lambda}}\ft F_2^{\lambda}\rp.
 \eea
For $n=-2$ we obtain
 \bea
 \frac{v_F^\lambda}{2}(\partial_x +i\partial_y)\ft F_{-1}^\lambda+ i 2 \omega_c \ft F_{-2}^\lambda &=& -\gamma_{ee}^\lambda \ft F_{-2}^\lambda -\gamma_d\lp \ft F_{-2}^\lambda-\frac{v_F^\lambda}{v_F^{\bar \lambda}} \ft F_{-2}^{\lambda '}\rp, \nn
 \frac{v_F^{\bar \lambda}}{2}(\partial_x +i\partial_y)\ft F_{-1}^{\bar \lambda}+ i 2 \omega_c \ft F_{-2}^{\bar \lambda} &=& -\gamma_{ee}^{\bar \lambda} \ft F_{-2}^{\bar \lambda} -\gamma_d\lp \ft F_{-2}^{\bar \lambda}-\frac{v_F^{\bar \lambda}}{v_F^{\lambda}} \ft F_{-2}^{\lambda }\rp.
 \eea
We can relate $\ft F_{\pm 1}^\lambda$ and $\ft F_{\pm 2}^\lambda$ in a matrix form as:
 \be
 \frac{\lb \partial_x\mp \partial_y\rb}{2}
 \begin{pmatrix}
   v_F^\lambda \ft F_{\pm 1}^\lambda\\
   v_F^{\bar \lambda} \ft F_{\pm 1}^{\bar \lambda}
 \end{pmatrix}
 =
 \begin {pmatrix}
   -\gamma_{ee}^\lambda -\gamma_d^\lambda\pm i 2\omega_c & \frac{v_F^\lambda}{v_F^{\bar \lambda}} \gamma_d^\lambda\\
   \frac{v_F^{\bar \lambda}}{v_F^{\lambda}}\gamma_d^{\bar \lambda} & -\gamma_{ee}^{\bar \lambda} -\gamma_d^{\bar \lambda} \pm i2 \omega_c
   \end{pmatrix}
 \begin{pmatrix}
  \ft F_{\pm 2}^\lambda\\
  \ft F_{\pm 2}^{\bar \lambda}
 \end{pmatrix}.
 \ee
In principle we can invert the matrix to get
 \be
 -\bar{D}^\pm \frac{\lb \partial_x \mp i\partial_y\rb}{2}
 \begin{pmatrix}
 v_F^\lambda \ft F_{\pm 1}^\lambda\\
 v_F^{\bar \lambda} \ft F_{\pm 1}^{\bar \lambda}
 \end{pmatrix}
 =
 \begin{pmatrix}
  \ft F_{\pm 2}^\lambda\\
  \ft F_{\pm 2}^{\bar \lambda}
 \end{pmatrix},
 \ee
where
 \bea
 \bar{D}^\pm& =& \mathcal{K}
 \begin{pmatrix}
 \gamma_d^{\bar \lambda}+\gamma_{ee}^{\bar \lambda}\mp 2i\omega_c & \displaystyle \frac{v_F^\lambda}{v_F^{\bar \lambda}}\gamma_d^\lambda\\
 \displaystyle \frac{v_F^{\bar \lambda}}{v_F^\lambda}\gamma_d^{\bar \lambda} & \gamma_d^\lambda +\gamma_{ee}^\lambda \mp 2 i \omega_c)
 \end{pmatrix},\\
 \mathcal{K} &=&  \frac{1}{\gamma_{ee}^\lambda\gamma_{ee}^{\bar \lambda}+\gamma_{d}^\lambda\gamma_{ee}^{\bar \lambda}+\gamma_{ee}^\lambda\gamma_{d}^{\bar \lambda}-4\omega_c^2 \mp 2i\omega_c(\gamma_d^\lambda+\gamma_d^{\bar \lambda}+\gamma_{ee}^\lambda+\gamma_{ee}^{\bar \lambda})}.\nonumber
 \eea
Separating the real and imaginary parts leads to
 \be
 -
 \lb
 \begin{pmatrix}
   R_{\lambda\lambda} &R_{\lambda \bar \lambda}\\
   R_{\bar \lambda\lambda} & R_{\bar \lambda\bar \lambda}
 \end{pmatrix}
 \pm i
  \begin{pmatrix}
   I_{\lambda\lambda} &I_{\lambda \bar \lambda}\\
   I_{\bar \lambda\lambda} & I_{\bar \lambda\bar \lambda}
  \end{pmatrix}
  \rb
  \frac{\lb \partial_x\mp \partial_y\rb}{2}
 \begin{pmatrix}
   v_F^\lambda \ft F_{\pm 1}^\lambda\\
   v_F^{\bar \lambda} \ft F_{\pm 1}^{\bar \lambda}
 \end{pmatrix}
 =
 \begin{pmatrix}
  \ft F_{\pm 2}^\lambda\\
  \ft F_{\pm 2}^{\bar \lambda}
 \end{pmatrix}.
 \ee
From this equation, we can see that the Coulomb drag affects the stress tensor and thus the viscosity. $\nabla \bar {\vec T}$ can be written as
\bea
-\frac{1}{m\bar n_\lambda}\lb \partial_x\bar T_{xx}^\lambda+\partial_y\bar T_{xy}^\lambda \rb&=&\frac{(v_F^{\lambda})^2}{4}\lb R_{\lambda\lambda}\nabla^2 u_x^\lambda + R_{\lambda\bar \lambda} \frac{\bar{n}^{\bar \lambda}}{\bar{n}^{\lambda}} \nabla^2u_x^{\bar \lambda} +I_{\lambda\lambda} \nabla^2u_y^\lambda  +I_{\lambda\bar \lambda}\frac{\bar{n}^{\bar \lambda}}{\bar{n}^{\lambda}}\nabla^2u_y^{\bar \lambda}\rb+\mathcal{O} (\partial_x \ft F_0^\lambda),\\
 -\frac{1}{m\bar n_\lambda}\lb\partial_x\bar T_{yx}^\lambda +\partial_y \bar T_{yy}^\lambda\rb&=&\frac{(v_F^{\lambda})^2}{4}\lb R_{\lambda\lambda}\nabla^2 u_y^\lambda  +R_{\lambda\bar \lambda}\frac{\bar{n}^{\bar \lambda}}{\bar{n}^{\lambda}} \nabla^2u_y^{\bar \lambda}-I_{\lambda\lambda} \nabla^2 u_x^\lambda -I_{\lambda\bar \lambda}\frac{\bar{n}^{\bar \lambda}}{\bar{n}^{\lambda}}\nabla^2u_x^{\bar \lambda}\rb+\mathcal{O} (\partial_y \ft F_0^\lambda),
 \eea
Using this relation, we write the Navier-Stokes equation explicitly as
\begin{align}
 \partial_t\vec u^\lambda+\gamma_d^{\lambda} (\vec u^\lambda-\vec u^{\bar \lambda}) = \nu^\lambda \nabla^2 \vec u^\lambda +\nu_H^{\lambda} \nabla^2\lp \vec u^\lambda \times \hat{\vec z} \rp
 +\nu_d^{\lambda} \nabla^2 \vec u^{\bar \lambda} +\nu_{dH}^{\lambda} \nabla^2\lp \vec u^{\bar \lambda} \times \hat{\vec z} \rp
+ \frac{e}{m} \boldsymbol{\mathcal E}^\lambda + \omega_c \vec u^\lambda \times \hat{\vec z},
\end{align}
where
 \bea
 \nu^\lambda &=& \frac{(v_F^\lambda)^2 R_{\lambda\lambda}}{4},\\
 \nu_d^\lambda &=& \frac{(v_F^\lambda)^2 R_{\lambda\bar \lambda}}{4}\frac{\bar n_{\bar \lambda}}{\bar n_{\lambda}},\\
 \nu_H^\lambda &=& \frac{(v_F^\lambda)^2 I_{\lambda\lambda}}{4},\\
 \nu_{dH}^\lambda &=& \frac{(v_F^\lambda)^2 I_{\lambda\bar \lambda}}{4}\frac{\bar n_{\bar \lambda}}{\bar n_\lambda},
 \eea
and  $\boldsymbol{\mathcal E}^\lambda= \vec E^\lambda + \frac{\alpha m}{e} \nabla P^\lambda=-\nabla \varphi^\lambda$ is the total electric field and gradient of pressure $\nabla P^\lambda =(\bar {n}_\lambda/g_F)\nabla n(\vec r,t)$. Now we write $\bar n_\one/\bar n_2=r$. In case of a parabolic band, this means that $v_F^\one/v_F^\two=\sqrt{r}$. For $\gamma_{ee}^\lambda\propto 1/n_\lambda$, we have thus $\gamma_{ee}^\one/\gamma_{ee}^\two=1/r$ and $\gamma_d^\lambda\propto \bar n_{\bar \lambda}$ means that $\gamma_d^\one/\gamma_d^\two=1/r$. For further simplification we drop the superscript ``$\one$'' and use $v_F^\one=v_F$ and $\gamma_{ee,d}^\one=\gamma_{ee,d}$.  We write the four viscosities in two layers explicitly as:
 \bea
 \nu^\one&=&\frac{v_F^2}{4\gamma_{ee}}\frac{r^2(1+3\Gamma_d+2\Gamma_d^2)+4\tilde\omega_c^2(1+\Gamma_d)}{8r\Gamma_d^2\tilde \omega_c^2+4\tilde\omega_c^2(1+2\Gamma_d+\Gamma_d^2+4\tilde\omega_c^2)+r^2(1+4\tilde\omega_c^2+4\Gamma_d^2(1+\tilde\omega_c^2)+\Gamma_d(4+8\tilde\omega_c^2))}\nn
 \nu_d^\one&=&\frac{v_F^2}{4\gamma_{ee}r}\frac{\sqrt{r}\Gamma_d(r+2r\Gamma_d-4\tilde\omega_c^2)}{8r\Gamma_d^2\tilde \omega_c^2+4\tilde\omega_c^2(1+2\Gamma_d+\Gamma_d^2+4\tilde\omega_c^2)+r^2(1+4\tilde\omega_c^2+4\Gamma_d^2(1+\tilde\omega_c^2)+\Gamma_d(4+8\tilde\omega_c^2))}\nn
 \nu_H^\one&=&\frac{v_F^2}{4\gamma_{ee}}\frac{2\tilde\omega_c(r\Gamma_d^2+r^2(1+\Gamma_d)^2+4\tilde\omega_c^2)}{8r\Gamma_d^2\tilde \omega_c^2+4\tilde\omega_c^2(1+2\Gamma_d+\Gamma_d^2+4\tilde\omega_c^2)+r^2(1+4\tilde\omega_c^2+4\Gamma_d^2(1+\tilde\omega_c^2)+\Gamma_d(4+8\tilde\omega_c^2))}\nn
 \nu_{dH}^\one&=&\frac{v_F^2}{4\gamma_{ee}r}\frac{2\sqrt{r}\tilde\omega_c\Gamma_d(1+r)(1+\Gamma_d)}{8r\Gamma_d^2\tilde \omega_c^2+4\tilde\omega_c^2(1+2\Gamma_d+\Gamma_d^2+4\tilde\omega_c^2)+r^2(1+4\tilde\omega_c^2+4\Gamma_d^2(1+\tilde\omega_c^2)+\Gamma_d(4+8\tilde\omega_c^2))}\nn
 \nu^\two&=&\frac{v_F^2}{4\gamma_{ee}r}\frac{r(1+3\Gamma_d+2\Gamma_d^2+4\tilde\omega_c^2(1+\Gamma_d))}{8r\Gamma_d^2\tilde \omega_c^2+4\tilde\omega_c^2(1+2\Gamma_d+\Gamma_d^2+4\tilde\omega_c^2)+r^2(1+4\tilde\omega_c^2+4\Gamma_d^2(1+\tilde\omega_c^2)+\Gamma_d(4+8\tilde\omega_c^2))}\nn
 \nu_d^\two&=&\frac{v_F^2}{4\gamma_{ee}}\frac{\sqrt{r}\Gamma_d(r+2r\Gamma_d-4\tilde\omega_c^2)}{8r\Gamma_d^2\tilde \omega_c^2+4\tilde\omega_c^2(1+2\Gamma_d+\Gamma_d^2+4\tilde\omega_c^2)+r^2(1+4\tilde\omega_c^2+4\Gamma_d^2(1+\tilde\omega_c^2)+\Gamma_d(4+8\tilde\omega_c^2))}\nn
 \nu_H^\two&=&\frac{v_F^2}{4\gamma_{ee}r}\frac{2\tilde\omega_c(1+2\Gamma_d+(1+r)\Gamma_d^2+4\tilde\omega_c^2)}{8r\Gamma_d^2\tilde \omega_c^2+4\tilde\omega_c^2(1+2\Gamma_d+\Gamma_d^2+4\tilde\omega_c^2)+r^2(1+4\tilde\omega_c^2+4\Gamma_d^2(1+\tilde\omega_c^2)+\Gamma_d(4+8\tilde\omega_c^2))}\nn
 \nu_{dH}^\two&=&\frac{v_F^2}{4\gamma_{ee}}\frac{2\sqrt{r}\tilde\omega_c\Gamma_d(1+r)(1+\Gamma_d)}{8r\Gamma_d^2\tilde \omega_c^2+4\tilde\omega_c^2(1+2\Gamma_d+\Gamma_d^2+4\tilde\omega_c^2)+r^2(1+4\tilde\omega_c^2+4\Gamma_d^2(1+\tilde\omega_c^2)+\Gamma_d(4+8\tilde\omega_c^2))},\nonumber
 \eea
 where $\Gamma_d=\gamma_d/\gamma_{ee}$ and $\tilde \omega_c=\omega_c/\gamma_{ee}$.

 In the limit of small $\tilde \omega_c$ we obtain
 \begin{align}
 \nu^\one  &=\nu_0 \frac{1+\Gamma_d}{1+2\Gamma_d}=\nu, &
 \nu_d^\one &= \frac{\nu_0}{r}\frac{\Gamma_d}{\sqrt{r}(1+2\Gamma_d))} = \frac{\nu_d}{r\sqrt{r}}\label{eq:nu1},\\
 \nu^\two &= \frac{\nu_0}{r}\frac{1+\Gamma_d}{r(1+2\Gamma_d)}=\frac{\nu}{r^2}, &
 \nu_d^\two &= \nu_0 \frac{\Gamma_d}{\sqrt{r}(1+2\Gamma_d)}=\frac{\nu_d}{\sqrt{r}},\\
 \nu_H^\one &=\frac{ 2\nu_0 \tilde \omega_c}{r} \frac{\Gamma_d^2 + r(1+\Gamma_d)^2}{(1+2\Gamma_d)^2},&
 \nu_{dH}^\one &= \frac{2\nu_0\tilde\omega_c}{r^{5/2}}\frac{2 (1+r) \Gamma_d(1+\Gamma_d)}{(1+2\Gamma_d)^2},\\
 \nu_H^\two &= \frac{2\nu_0 \tilde\omega_c }{r^3} \frac{1+2\Gamma_d+(1+r)\Gamma_d^2}{(1+2 \Gamma_d)^2},&
 \nu_{dH}^\two &=\frac{ 2\nu_0 \tilde\omega_c}{r^{3/2}}\frac{2(1+r)\Gamma_d(1+\Gamma_d)}{ (1+2\Gamma_d)^2} \label{eq:nu2},
 \end{align}
 where $\nu_0=\frac{v_F^2}{4\gamma_{ee}}$.
 In the simplest case, we can assume two identical liquids in the top and bottom layers ($r=1$), in which case we obtain identical viscosities in the two layers,
 \bea
 \nu &=& \frac{ v_F^2}{4}\frac{(\gamma_{d}+\gamma_{ee})[\gamma_{ee}(2\gamma_d+\gamma_{ee})+4\omega_c^2]}{(\gamma_{ee}^2+4\omega_c^2)[(\gamma_{ee}+2\gamma_d)^2+4\omega_c^2]},\nn
 \nu_H &=& \frac{ v_F^2}{4}\frac{2\omega_c[\gamma_{ee}(\gamma_{ee}+2\gamma_d)+2\gamma_d^2+4\omega_c^2])}{(\gamma_{ee}^2+4\omega_c^2)[(\gamma_{ee}+2\gamma_d)^2+4\omega_c^2]},\\
 \nu_{d} &=& \frac{ v_F^2}{4}\frac{\gamma_{d}[\gamma_{ee}(2\gamma_d+\gamma_{ee})-4\omega_c^2]}{(\gamma_{ee}^2+4\omega_c^2)[(\gamma_{ee}+2\gamma_d)^2+4\omega_c^2]},\nn
 \nu_{dH} &=& \frac{ v_F^2}{4}\frac{4\gamma_{d}\omega_c(\gamma_d+\gamma_{ee})}{(\gamma_{ee}^2+4\omega_c^2)[(\gamma_{ee}+2\gamma_d)^2+4\omega_c^2]}.\nonumber
 \eea

 \section{Coulomb drag scattering integral} \label{sec:collision}
 We consider the Boltzmann equation including only the drag scattering integral,
 \be
 \partial_t f^\lambda+\vec v_\vec p^\lambda \cdot \partial_{\vec r} f^\lambda +\lp e \vec E^\lambda+ \frac{e}{c}\vec v_{\vec p}^\lambda \times \vec B \rp \cdot \partial_\vec p f^\lambda =  S_{\lambda,\bar \lambda}^d(\vec p),\label{eq:kineticd}
 \ee
where the scattering integral is given by
 \bea
 S_{\lambda,\bar \lambda}^d(\vec p) &=&-\int \frac{d^2\vec p_2}{(2\pi \hbar)^2} \int \frac{d^2\vec q}{(2\pi\hbar)^2} \delta(\eps_\vec p^\lambda+\eps_\vec {p_2}^{\bar \lambda}-\eps_\vec{p+q}^\lambda-\eps_\vec{p_2-q}^{\bar \lambda}) |V_\vec {p,p_2,q}^{\lambda,\bar \lambda}|^2\nn
&& \times \lb f^\lambda_\vec p f^{\bar \lambda}_\vec {p_2}\lp1-f^{\lambda}_\vec{p+q} \rp \lp 1- f^{\bar \lambda}_\vec{p_2-q}\rp - \lp 1-f^\lambda_\vec p \rp \lp 1- f^{\bar \lambda}_\vec {p_2}\rp f^{\lambda}_\vec{p+q}  f^{\bar \lambda}_\vec{p_2-q}\rb.\label{eq:sd}
 \eea
 In hydrodynamics the local equilibrium distribution is given by:
 \be
 f^\lambda_\vec p = \frac{1}{1+ \exp(\beta (\eps^\lambda_\vec p-\mu^\lambda -\vec p \cdot \vec u^\lambda))}
 \ee
 We linearize the collision integral~\eqref{eq:sd} by expanding $f^\lambda_\vec p$ to linear order in $u$ as
 \bea
 f^\lambda_\vec p &=& f^{\zero\lambda}_\vec p +\delta f =f^{\zero\lambda}_\vec p   - \vec p \cdot \vec u \frac{\partial f^{\zero\lambda}_\vec p  }{\partial \eps}\nn
 &=& f^{\zero\lambda}_\vec p - (\vec p \cdot \vec u) \beta f^{\zero\lambda}_\vec p(1-f^{\zero\lambda}_\vec p),
 \eea
 where $f^{\zero \lambda}=\{1+\exp [\beta (\eps_\vec p^\lambda -\mu^\lambda)]\}^{-1}$.
 Linearizing the collision integral means retaining the linear order of the distribution function products,
 \bea
 f_1 f_2\lp1-f_3 \rp \lp 1- f_4\rp &=&   (f_1^\zero +\delta f_1)( f_2^\zero+\delta f_2)\lp 1-(f_3^\zero+\delta f_3) \rp \lp 1- (f_4^\zero +\delta f_4)\rp\nn
 &\approx& f_1^\zero f_2^\zero\lp1-f_3^\zero \rp \lp 1- f_4^\zero\rp + \delta f_1 f_2^\zero\lp1-f_3^\zero \rp \lp 1- f_4^\zero\rp\nn
 && + f_1^\zero \delta f_2\lp1-f_3^\zero \rp \lp 1- f_4^\zero\rp  + f_1^\zero  f_2^\zero(-\delta f_3) \lp 1- f_4^\zero\rp\nn
 && + f_1^\zero  f_2^\zero\lp1-f_3^\zero \rp (-\delta f_4^\zero)
 \eea
 The same way, we get:
 \bea
 (1-f_1) (1-f_2) f_3  f_4 &\approx& (1-f_1^\zero)(1- f_2^\zero) f_3^\zero  f_4^\zero +(-\delta f_1) (1-f_2^\zero)f_3^\zero f_4^\zero\nn
 && + (1- f_1^\zero) (-\delta f_2) f_3^\zero  f_4^\zero  + (1-f_1^\zero)(1-  f_2^\zero)(\delta f_3) f_4^\zero\nn
 && + (1-f_1^\zero)(1-  f_2^\zero)f_3^\zero  (\delta f_4^\zero)
 \eea
 Thus:
 \bea
\lb  f_1 f_2\lp1-f_3 \rp \lp 1- f_4\rp - (1-f_1) (1-f_2) f_3  f_4\rb
 &\approx&
 \delta f_1 \lb f_2^\zero \lp 1- f_3^\zero \rp \lp 1-f_4^\zero\rp + (1-f_2^\zero) f_3^\zero f_4^\zero\rb\nn
 &&+ \delta f_2 \lb f_1^\zero \lp 1- f_3^\zero \rp \lp 1-f_4^\zero\rp + (1-f_1^\zero) f_3^\zero f_4^\zero\rb \nn
 &&-\delta f_3 \lb f_1^\zero f_2^\zero \lp 1-f_4^\zero\rp + \lp 1-f_1^\zero\rp \lp 1-f_2^\zero \rp f_4^\zero    \rb \nn
 &&- \delta f_4 \lb f_1^\zero f_2^\zero \lp 1-f_3^\zero\rp + \lp 1-f_1^\zero\rp \lp 1-f_2^\zero \rp f_3^\zero    \rb ,\nn
 &\approx&
 f_1^\zero f_2^\zero\lp1-f_3^\zero \rp \lp 1- f_4^\zero\rp \lb -h_1 -h_2 + h_3 + h_4\rb
 \eea
 where we have used $  f_1^\zero f_2^\zero(1-f_3^\zero )( 1- f_4^\zero) - (1-f_1^\zero) (1-f_2^\zero) f_3^\zero  f_4^\zero=0$ from the energy and number conservations and $\delta f_1 = -h_1 f_1^\zero (1-f_1^\zero) $, where $h_1 = \beta \vec p_1 \cdot \vec u_1$.
 The linearized collision integral reads
 \bea
 S_{\lambda,\bar \lambda}^d(\vec p) &\approx&-\int \frac{d^2\vec p_2}{(2\pi \hbar)^2} \int \frac{d^2\vec q}{(2\pi\hbar)^2} \delta(\eps_\vec p^\lambda+\eps_\vec {p_2}^{\bar \lambda}-\eps_\vec{p+q}^\lambda-\eps_\vec{p_2-q}^{\bar \lambda}) |V_\vec {p,p_2,q}^{\lambda,\bar \lambda}|^2 \beta\nn
 && \times \lb f^{\zero \lambda}_\vec p f^{\zero \bar \lambda}_\vec {p_2}\lp1-f^{\zero \lambda}_\vec{p+q} \rp \lp 1- f^{\zero \bar \lambda}_\vec{p_2-q}\rp\rb \lb - \vec p \cdot \vec u^\lambda - \vec p_2 \cdot \vec u^{\bar \lambda} + \lp\vec p + \vec q \rp \cdot \vec u ^\lambda + \lp\vec p_2 - \vec q \rp \cdot \vec u ^{\bar \lambda }   \rb.\nn
 &=&-(\vec u^\lambda -\vec u^{\bar \lambda})\cdot \int \frac{d^2\vec p_2}{(2\pi \hbar)^2} \int \frac{d^2\vec q}{(2\pi\hbar)^2}\  \vec q\  \delta(\eps_\vec p^\lambda+\eps_\vec {p_2}^{\bar \lambda}-\eps_\vec{p+q}^\lambda-\eps_\vec{p_2-q}^{\bar \lambda}) |V_\vec {p,p_2,q}^{\lambda,\bar \lambda}|^2 \beta\nn
 && \times \lb f^{\zero \lambda}_\vec p f^{\zero \bar \lambda}_\vec {p_2}\lp1-f^{\zero \lambda}_\vec{p+q} \rp \lp 1- f^{\zero \bar \lambda}_\vec{p_2-q}\rp\rb\label{eq:sdlin}.
 \eea
 
 \begin{figure}[t]
\centering
 \includegraphics[height=6cm]{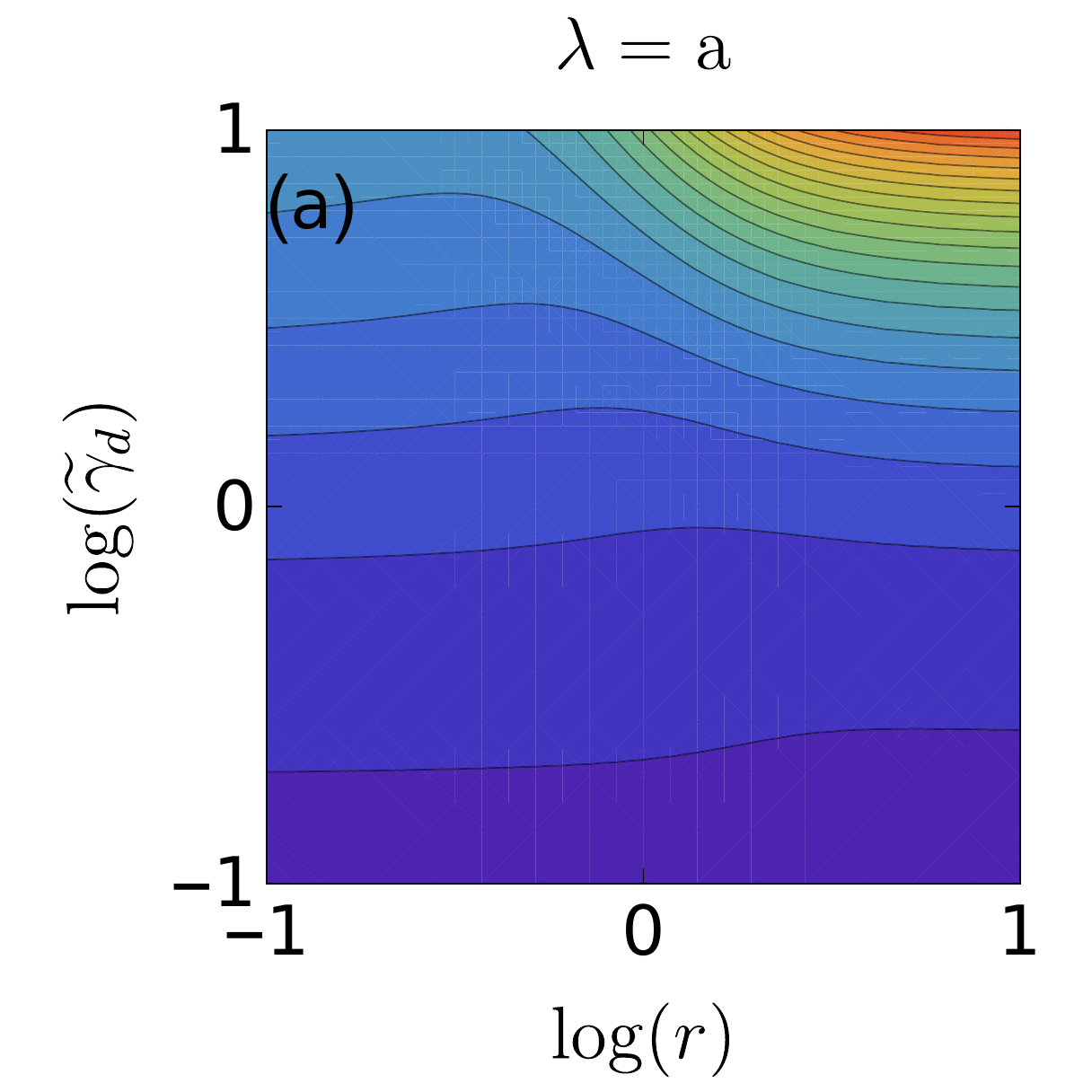}
 \includegraphics[height=6cm]{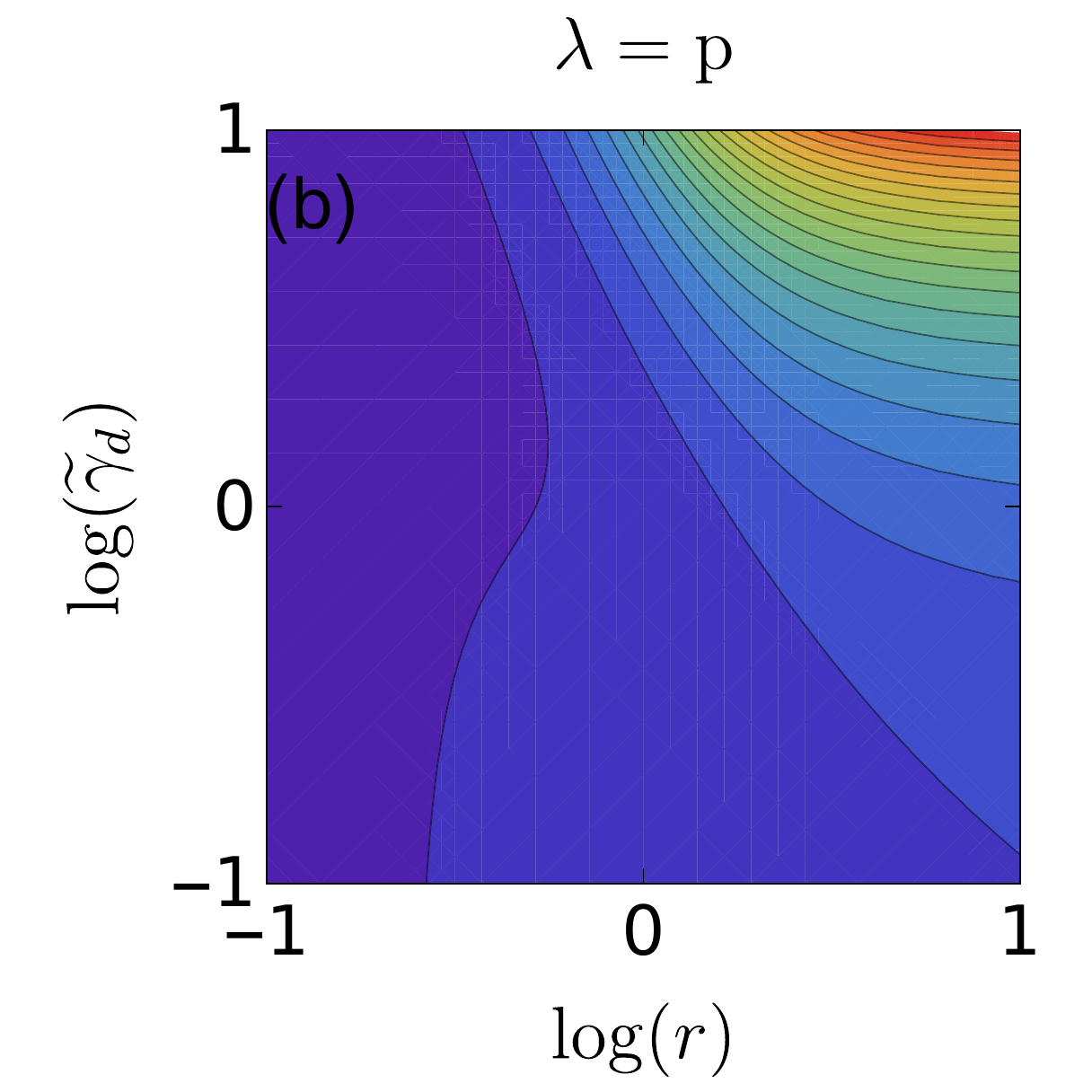}
 \includegraphics[trim=0cm 3cm 0cm 3cm,clip,height=8cm]{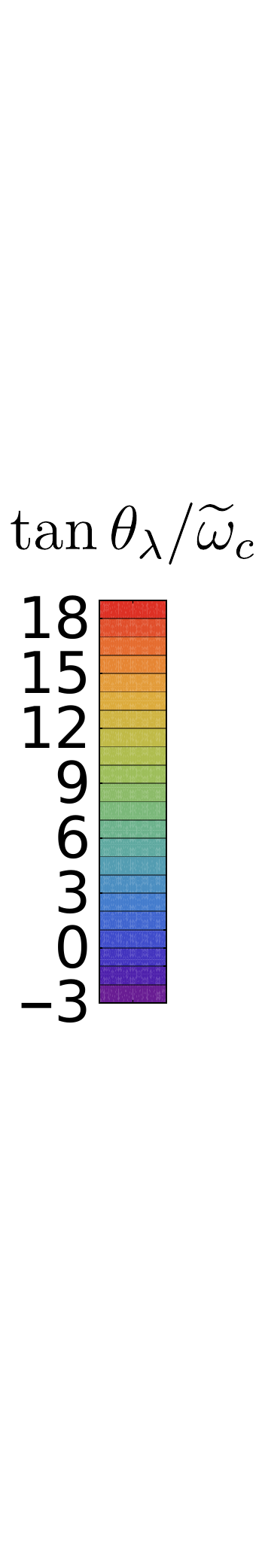}
\caption{\label{fig:2dHangle} (a) Hall angles at the first layer $\tan\theta_\one/\tilde{\omega}_c$ and  (b) at the the second layer $\tan\theta_\two/\tilde{\omega}_c$  as functions of $r=\bar n_\one/\bar n_\two$ and $\tilde{\gamma}_d$. }
\end{figure}

 Note that $\int d^2\vec p S^d_{\lambda,\bar \lambda} (\vec p) =0$ because $\langle \cos (\theta_\vec q)\rangle_\theta = 0$. On the other hand, $\int d^2\vec p\ \vec p\ S^d_{\lambda,\bar \lambda}(\vec p)$ is nonzero because $\langle \cos^2 (\theta_\vec q)\rangle_\theta = 1/2$.
Looking at the shape of $S_{\lambda,\bar \lambda}^d(\vec p)$ the collision integral should be proportional to $\bar n_\lambda \bar n_{\bar \lambda}$.
We multiply $\vec p^\lambda/m $ into Boltzmann equation~\eqref{eq:kineticd} and integrate over $\vec p$, we obtain
\be
\partial_t \vec j^\lambda +\frac{1}{m} \nabla \bar \Pi -\lp e\bar n ^\lambda E^\lambda +\frac{e}{c}\vec j^\lambda \times \vec B\rp = -\gamma_d \bar n_\lambda (\vec u^\lambda -\vec u^{\bar \lambda})
\ee
Writing down the Navier Stokes equation in terms of $\vec u$ we obtain:
\be
\partial_t \vec u^\lambda +\frac{1}{\bar n_\lambda m} \nabla \bar \Pi -\lp e E^\lambda +\frac{e}{c}\vec u^\lambda \times \vec B\rp = -\gamma_d^\lambda  (\vec u^\lambda -\vec u^{\bar \lambda})
\ee
where $\gamma_d^\lambda$ is proportional to $\bar n_{\bar \lambda}$.

\section{Hall dynamics and Hall angle}
Starting from Eq.~(7) in the main text, we will evaluate the Hall angle $\tan \theta_\lambda=E_y^\lambda/E_x$.
We express the equation in terms of dimensionless quantities to get
\bea
\frac{\nu_H^\lambda u_0}{w_h^2} \partial_{\tilde y}^2 \tilde u^\lambda  + \frac{\nu_{dH}^\lambda u_0}{w_h^2} \partial_{\tilde y}^2 \tilde u^{\bar \lambda}+ \omega_c u_0 u^\lambda&=&\frac{e}{m}E_y^\lambda ,\quad u_0 =\frac{eE_x w_h^2}{m\nu} \nn
\frac{\nu_H^\lambda \gamma_{ee}}{\nu \omega_c} \partial_{\tilde y}^2 \tilde u^\lambda +  \frac{\nu_{dH}^\lambda \gamma_{ee}}{\nu \omega_c} \partial_{\tilde y}^2 \tilde u^{\bar \lambda} + \frac{\gamma_{ee} w_h^2}{\nu} \tilde u^\lambda &=&\frac{\gamma_{ee}}{\omega_c} \tan \theta_\lambda\nn
\tilde \nu_H^\lambda \partial_{\tilde y}^2 \tilde u^\lambda +  \tilde \nu_{dH}^\lambda \partial_{\tilde y}^2 \tilde u^{\bar \lambda} + \mathcal{R} \tilde u^\lambda &=& \frac{\tan\theta_\lambda}{\tilde \omega_c}
\eea
From Eqs.~\eqref{eq:nu1}--\eqref{eq:nu2}, we obtain
\bea
\tilde \nu_H^\one = \frac{2}{r}\frac{\Gamma_d^2 + r (1+\Gamma_d)^2}{(1+\Gamma_d)(1+2\Gamma_d)}& , & \tilde \nu_{dH}^\one = \frac{2}{r^{5/2}}\frac{2(1+ r) \Gamma_d }{1+2\Gamma_d}\nn
\tilde \nu_H^\two = \frac{2}{r^3}\frac{1+2\Gamma_d+(1+r)\Gamma_d^2 }{(1+\Gamma_d)(1+2\Gamma_d)}& , & \tilde \nu_{dH}^\two = \frac{2}{r^{3/2}}\frac{2(1+ r) \Gamma_d }{1+2\Gamma_d}.
\eea
The Reynolds number $\displaystyle \mathcal{R}= \frac{\gamma_{ee}^2 w_h^2}{v_F^2}\frac{1+2\Gamma_d}{1+\Gamma_d}$ is closely related to $\displaystyle \tilde \gamma_d=\frac{\gamma_{ee}\gamma_d w_h^2}{v_F^2}\frac{1+2\Gamma_d}{1+\Gamma_d}=\Gamma_d \mathcal{R}$.

\end{widetext}

\bibliography{refs}

\end{document}